\documentclass[runningheads,11pt]{llncs}

\usepackage[a4paper,margin=2.5cm]{geometry}
\usepackage{hyperref}
\usepackage[x11names]{xcolor}
\hypersetup{
colorlinks=true,
citecolor=blue
}
\usepackage{orcidlink}
\usepackage{amssymb,amsmath}
\usepackage{mathtools}
\usepackage{xspace}
\usepackage[percent]{overpic}
\usepackage{graphicx}
\usepackage{comment,multirow}
\usepackage[figure,tworuled,linesnumbered,vlined]{algorithm2e} 
\usepackage{tikz}

\definecolor{ser}{rgb}{0.95, 0.1, 0.1}
\definecolor{alf}{HTML}{0030f3}
\definecolor{ale}{HTML}{1d8348}
\definecolor{gab}{HTML}{dd13F8}

\newcommand{\nb}[2]{
\fcolorbox{gray}{yellow}{\bfseries\sffamily\scriptsize#1}
{\sf\small$\blacktriangleright$\textit{#2}$\blacktriangleleft$}
}
\newcommand{\alf}[1]{\textcolor{alf}{\nb{Alf}{#1}}}

\newcommand{\Look}{{\tt Look}\xspace}

\newcommand{\LCM}{{\tt LCM}\xspace}

\newcommand{\vet}{\mathcal{VET}}

\newcommand{\A}{\mathcal{A}} 
\newcommand{\Ex}{\mathbb{E}} 
\newcommand{\rr}{$\mathsf{RR}$}

\newcommand{\pre}{\mathtt{pre}}

\newcommand{\mbr}{\mathit{mbr}}

\newcommand{\DMA}{\mathtt{DMA}}
\newcommand{\Ust}{U_\mathbf{ST}}
\newcommand{\Uh}{U_\mathbf{H}}

\newcommand{\algosquare}{\mathcal{A}_\mathbf{ST}}
\newcommand{\algohyp}{\mathcal{A}_\mathbf{H}}

\newcommand{\ctype}[2]{(#1\!\times\! #2)}

\newcommand{\false}{\mathrm{false}}

\newcommand{\block}[1]{\smallskip\noindent{\textbf{#1.}}~}

\newcommand{\gath}{\sc{Gathering}\xspace}

\newcommand{\mbh}{\mathit{mbh}}

\SetKwComment{tcp}{$\#$ }{}%

\SetCommentSty{mycommfont}

\begin{document}
\title{Gathering in Vertex- and Edge-Transitive Graphs without Multiplicity Detection under Round Robin} 

\titlerunning{Gathering in graphs under round robin}
\author{%
Serafino Cicerone\inst{1}
\and
Alessia {Di Fonso}\inst{1}
\and
Gabriele {Di Stefano}\inst{1}
\and\\
Alfredo Navarra\inst{2}
}
\authorrunning{S. Cicerone et al.}
%
\institute{
        Università degli Studi dell'Aquila, I-67100 
        L'Aquila, Italy. \\
\email{serafino.cicerone@univaq.it}, \email{alessia.difonso@univaq.it}, \email{gabriele.distefano@univaq.it}
\and 
        Università degli Studi di Perugia I-06123 
        Perugia, Italy.
\email{alfredo.navarra@unipg.it}
}

\maketitle

\begin{abstract}
In the field of swarm robotics, one of the most studied problem is {\gath}. It asks for a distributed algorithm that brings the robots to a common location, not known in advance. We consider the case of robots constrained to move along the edges of a graph under the well-known $\mathcal{OBLOT}$ model. Gathering is then accomplished once all the robots occupy a same vertex. Differently from classical settings, we assume: i) the initial configuration may contain \emph{multiplicities}, i.e. more than one robot may occupy the same vertex; ii) robots cannot detect multiplicities; iii) robots move along the edges of vertex- and edge-transitive graphs, i.e. graphs where all the vertices (and the edges, resp.) belong to a same class of equivalence. To balance somehow such a `hostile' setting, as a scheduler for the activation of the robots, we consider the \emph{round-robin}, where robots are cyclically activated one at a time.

We provide some basic impossibility results and we design two different algorithms approaching the {\gath} for robots moving on two specific topologies belonging to edge- and vertex-transitive graphs: infinite grids and hypercubes. The two algorithms are both time-optimal and heavily exploit the properties of the underlying topologies. Because of this, we conjecture that no general algorithm can exist for all the solvable cases.
\end{abstract}


\section{Introduction}
In the field of distributed computing, 
swarm robotics is attracting more and more researchers. Robots are typically modeled as autonomous mobile units that, despite acting independently, are capable of exhibiting collective behavior to solve a shared task. Modeling robots from a theoretical perspective usually results in minimizing available capabilities in order to understand what can be computed while maximizing flexibility and robustness to faults. In this respect, a particularly prominent model is the so-called $\mathcal{OBLOT}$~\cite{FPS-macbook19,FPS12}. Each robot alternates between inactive periods and active {\tt Look-Compute-Move} ({\tt LCM}) cycles: when active, a robot acquires a snapshot of its surroundings ({\tt Look}), computes its next move according to a provided algorithm ({\tt Compute}), and then proceeds to reach the selected destination ({\tt Move}).

Among the tasks studied in $\mathcal{OBLOT}$, {\gath} is certainly one of the most popular. Robots are required to reach a common, but unspecified, place from where they do not move anymore. 
In the case of robots freely moving in the Euclidean plane, the {\gath} problem has been fully characterized (see~\cite{CFPS12,SY99}). In fact, it is always solvable for synchronous robots, whereas in the asynchronous case it is unsolvable when just two robots are considered.

When robots are constrained to move along the edges of a graph, more investigation is required. In fact, apart for some impossibility results or basic conditions that guarantee the resolution of the problem
provided in \cite{CDN18-book,CDN20a,DN17a}, most of the literature usually approaches specific topologies.
Studied topologies are: 
Trees \cite{DDKN12,DDN13}, Regular Bipartite graphs~\cite{GP13}, Finite Grids \cite{DDKN12}, 
Infinite Grids \cite{DN17}, 
Tori \cite{KLOTW21}, Oriented Hypercubes \cite{BKAS18}, Complete graphs~\cite{CDN19c,CDN20a}, 
Complete Bipartite graphs \cite{CDN19c,CDN20a}, Butterflies \cite{CDDN24,CiceroneFSN25_tcs}, 
and Rings \cite{BPT16,DDN14,DDN11,DDNNS15,DNN17,DDFN18,DN17a,IIKO13,KKN10,KMP08,NP25,OT12}.	

Most of such topologies are very symmetric, i.e., the vertices can be partitioned into a few classes of equivalence. Since robots have few topological properties to exploit, the design of a solving algorithm becomes more challenging.
In particular, \emph{vertex- and edge-transitive} graphs, i.e., graphs where all the vertices and all edges are topologically equivalent, turn to be very difficult to approach. Examples of such graphs are complete graphs, complete bipartite graphs $K_{n,n}$, rings, hypercubes or infinite grids.


Another crucial aspect of the cited works, applicable to both Euclidean and graphs, is the time scheduling under which the robots operate. Notably, the \emph{asynchronous} scheduler, where robots are activated independently of one another, often presents the greatest challenges. However, there are cases where asynchrony makes the problem unsolvable, whereas the \emph{synchronous} setting allows the development of non-trivial strategies, see, e.g.~\cite{CDN20a}.

In all the aforementioned settings, in the literature it is usually assumed that: 

\begin{description}
    \item[$A_1$:] robots are endowed with some kind of {\emph{multiplicity detection}}. With this, robots are able to recognize whether a vertex contains a {\emph{multiplicity}}, i.e., if two or more robots are located at the same vertex (not necessarily the exact number); 
    \item[$A_2$:] the \emph{starting} configuration, i.e., the first configuration ever seen by any robot,  does not contain multiplicities.     
\end{description}

In this paper, we remove assumptions $A_1$ and $A_2$, while as the time scheduler, we consider  {\emph{Round Robin}} (\rr). This is a specific type of synchronous scheduler, where $k$ robots are activated one at a time in a fair sequence. That is, in $k$ {\tt Look-Compute-Move} cycles, all robots are activated. This constitutes an \emph{epoch}. The order of activations is then repeated in each epoch. 

Certainly, dealing with {\rr} may seem much easier compared to other schedulers, especially the asynchronous one. Indeed, all the symmetries admitted by a given configuration are inherently broken each time by the single activated robot. 
On the contrary, the {\gath} problem becomes considerably more complicated without assumptions $A_1$ and $A_2$.

\block{Results}
For robots moving on non-vertex-transitive graphs (i.e., graphs where the vertices are partitioned into at least two different classes of equivalence), the {\gath} problem without assumptions $A_1$ and $A_2$ under {\rr} has been fully solved in~\cite{CDDN25-gathering-noVT,SSS_vertex_nonVT_RR}. For robots moving on rings (vertex- and edge-transitive topology), under the same assumptions, the problem has been fully characterized in~\cite{NP25}. In this paper, we keep investigating the {\gath} problem under the \rr\ scheduler, without relying on assumptions $A_1$ and $A_2$, and considering the whole family of vertex- and edge-transitive graphs.

The uniformity of the vertex and edge structure poses significant challenges in devising a general solution strategy. We begin by identifying certain instances where gathering is impossible. We then solve the {\gath} problem for specific topologies like infinite grids and hypercubes. The two algorithms $\algosquare$ and $\algohyp$ that we propose for these cases are time-optimal, but they rely heavily on specific characteristics of each topology. Our findings indicate that 
a single general strategy able to solve {\gath} is highly unlikely.

\block{Outline} In the next section, we recall the $\mathcal{OBLOT}$ model. In Section~\ref{sec:prob}, we formulate the {\gath} problem and  introduce some useful notation. We also summarize the adopted methodology for formally describe algorithms in the assumed robot model. 
In Section~\ref{sub:imposs}, we prove some impossibility results, and then we provide two distinct algorithms solving the {\gath} problem for robots moving on hypercubes (Section~\ref{sub:cubes}), and infinite grids (Section~\ref{sec:grids-tmp}) respectively. Finally, Section~\ref{sec:concl} contains concluding remarks and future work suggestions.

\section{Model}\label{sec:model}
We consider the standard  $\mathcal {OBLOT}$ model of 
distributed systems of autonomous mobile robots. In $\mathcal {OBLOT}$, the system is composed of a set 
 $\mathcal{R} = \{r_1, r_2, \dots, r_k\}$ of $k\ge 2$ computational {\emph robots} that  operate on a graph  $G$.
 Each vertex of $G$ is initially empty, occupied by one robot, or occupied by more than one robot (i.e., a \emph{multiplicity}; recall that we are not using assumptions $A_1$ and $A_2$ as described in the introduction).
 

Robots can be characterized 
  according to many different settings. In particular, they have the following basic properties: 
  
  \begin{itemize}
      \item \textbf{Anonymous:} they have no unique identifiers;
      \item \textbf{Autonomous:} they operate without a centralized control;
      \item \textbf{Dimensionless:} they are viewed as points, i.e., they have no volume nor occupancy restraints;
      \item \textbf{Disoriented:} they have no common sense of orientation;
      \item \textbf{Oblivious:} they have no memory of past events;
      \item \textbf{Homogeneous:} they all execute the same deterministic algorithm with no type of randomization admitted;
      \item \textbf{Silent:} they have no means of direct communication.
  \end{itemize}
  
 

Each robot in the system has sensory capabilities, allowing it to determine the location of
other robots in the graph, relative to its location. Each robot refers to a {\tt Local Reference System}
({\tt LRS}) that might differ from robot to robot. 
Each robot has a specific behavior described according to the sequence of the following four states: {\tt Wait}, {\tt Look}, {\tt Compute}, and {\tt Move}. Such a sequence defines the computational activation cycle (or simply a cycle) of a robot. More in detail:

\begin{enumerate}
    \item {\tt Wait:} the robot is in an idle state and cannot remain as such indefinitely;
    \item {\tt Look:} the robot obtains a global snapshot of the system containing the positions of the other robots with respect to its {\tt LRS}, by activating its sensors. Each robot is seen as a point in the graph occupying a vertex;
    \item {\tt Compute:} the robot executes a local computation according to a deterministic algorithm $\mathcal{A}$ (we also say that the robot executes 
    $\mathcal{A}$). This algorithm is the same for all the robots, and its result is the destination of the movement of the robot. Such a destination is either 
    the vertex where the robot is already located, or a neighboring vertex at one hop distance (i.e., only one edge per move can be traversed); 
    \item {\tt Move:} if the computed destination is a neighboring vertex $v$, the robot moves to $v$; otherwise, it executes a {\emph{nil}} 
    movement (i.e., it does not move).
\end{enumerate}

In the literature, the computational cycle is simply referred to as {\tt Look-Compute-Move} ({\tt LCM}) cycle, because when a robot is in the {\tt Wait} state, we say that it is 
\emph{inactive}. Thus, the {\tt LCM} cycle only refers to the \emph{active} states of a 
robot. 
It is also important to notice that since the robots are oblivious, without 
memory of past events, every decision they make during the {\tt Compute} phase is 
based on what they can determine during the current {\tt LCM} cycle. In 
particular, during the {\tt Look} phase, the robots take a snapshot of the system and 
they use it to elaborate the information, building what is called the \emph{view} of the 
robot. 
Regarding the {\tt Move} phase of the robots, the movements executed are always considered to be instantaneous. Thus, the robots are only able to perceive the 
other robots positioned on the vertices of the graph, never while moving. Regarding the 
position of a robot on a vertex, two or more robots may be located on 
the same vertex, i.e., they constitute a multiplicity.

Another important feature that can greatly affect the computational power of the robots is the \emph{time scheduler}. 
In this work, we consider the standard Round Robin (\rr): 
\begin{itemize}
    \item 
   Robots are activated one at time.
   If there are $k$ robots, then from round $1$ to round $k$, all the robots are activated. In the subsequent $k$ rounds,  all the robots are again activated in the same order. Each of those intervals of $k$ rounds is said to be an \emph{epoch}.
\end{itemize}

\section{Problem formulation and preliminary observations}\label{sec:prob}

The topology where robots are placed on is represented by a simple and connected graph $G=(V,E)$, with finite vertex set $V(G)=V$ and finite edge set $E(G)=E$ 
(the only exception is in Section~\ref{sec:grids-tmp} where we deal with infinite grids). 
The family of graphs which are both vertex- and edge-transitive is denoted as $\vet$. 
The cardinality of $V$ is represented as $|V|$ or $n$. $G[X]$ denotes the subgraph of $G$ induced by a subset of vertices $X\subset V$. We denote by $diam(G)$ the diameter of $G$, that is, the maximum distance between any pair of vertices of the graph. For each vertex $v\in V$, $N(v)$ is the set of neighboring vertices of $v$ and $N[v] = N(v)\cup \{v\}$.

A function $\lambda: V\to \{0,1\}$, from the set of vertices to the set $\{0,1\}$, indicates to the robots whether a vertex of $G$ is empty or occupied. Note that more than one robot may occupy the same vertex, but robots cannot perceive such information. We call $C=(G,\lambda)$ a \emph{configuration} whenever the actual number of robots is bounded and greater than zero.
%
A subset $V'\subseteq V$ is said to be \emph{occupied} if at least one of its elements is occupied, and \emph{unoccupied} otherwise.
%
We denote by $\Delta(C)$ the maximum distance among any pair of vertices that are occupied in $C$, and by $occ(C)=\sum_{v\in G} \lambda(v)$ the number of vertices that are occupied in $C$.

A configuration $C$ is \emph{final} if all the robots are on a single vertex (i.e., $\exists u\in V: \lambda(u)=1$ and $\lambda(v)=0,~\forall v\in V \setminus \{u\}$), i.e. $\Delta(C)= 0$.
Any configuration $C$ that is not final can be \emph{initial}.
Gathering can be formally defined as the problem of transforming an initial configuration into a final one. Throughout the paper, we assume that each initial configuration is composed of at least two robots occupying at least two vertices (otherwise, the problem is trivially solved). A \emph{gathering algorithm} for this problem is a deterministic distributed algorithm that brings the robots in the system to a final configuration in a finite number of \LCM-cycles from any given initial configuration, regardless of the adversary. %
Let $C(i)$ represent the configuration $C$ as observed at time instant $i$.
Formally, an algorithm $\A$ solves Gathering for an \emph{initial configuration} $C$ if,  for any execution $\Ex : C=C(0),C(1),C(2),\ldots$ of $\A$, there exists a time instant $t>0$ such that $C(t)$%
is final and no robots move after $t$, i.e., $C(t') = C(t)$ holds for all $t'\ge t$. 
Given an initial configuration $C=(G,\lambda)$, note that many different placements of robots correspond to $C$ due to possible multiplicities. If there exists a gathering algorithm for all the placements of robots corresponding to $C$, we say that $C$ is \emph{gatherable}; otherwise, we say that $C$ is \emph{ungatherable}.  
With respect to the number of epochs required to accomplish Gathering, we can state the following:

\begin{lemma}\label{lem:lb}
Let $C=(G,\lambda)$ be any initial configuration. Any solving algorithm for  {\gath} on $C$ under \rr\ requires $\Omega(diam(G))$ epochs. 
\end{lemma}
\begin{proof}
Let $r$ and $r'$ be two robots occupying two vertices that determine $\Delta(C)$. In order to solve the {\gath} problem, $r$ and $r'$ have to meet, eventually. The fastest way to do it is that they move toward each other along a shortest path. This requires $\Delta(C)/2$ moves for each robot and hence, due to the \rr\ scheduler, $\Delta(C)/2$ epochs.
As the maximum distance among robots in $C$ is at most $diam(G)$, by the generality of $C$, the claim holds.\qed
\end{proof}

\block{Our approach and methodology} 
The algorithms presented in this paper to solve the {\gath} problem on general graphs follow the methodology proposed in \cite{CDN21a}. Here, we briefly outline how a generic algorithm $\mathcal{A}$, intended to solve a problem $\mathcal{P}$, can be designed according to this approach.

In our operational model, robots have extremely limited capabilities. Therefore, it is often beneficial to decompose the main problem $\mathcal{P}$ into simpler sub-problems. Each sub-problem is associated with a ``task'' that can be executed by one or more robots. Let us denote these tasks as $T_1, T_2, \dots, T_q$. Among these tasks, at least one is designated as the \emph{terminal} task. This task means the main problem $\mathcal{P}$ has been solved and the robots need to recognize that no further action is required.


Since robots operate according to the {\tt LCM} cycle, they must identify the correct task to execute based on the configuration they observe during the {\tt Look} phase. This task recognition is triggered by associating a predicate $P_i$ with each task $T_i$. When a robot wakes up and detects that a predicate $P_i$ holds, it understands that it must execute task $T_i$.

More concretely, if the predicates are well defined, algorithm $\mathcal{A}$ can use them during the {\tt Compute} phase as follows: if a robot $r$ observes that predicate $P_i$ is true, then $r$ executes the corresponding move $m_i$ associated with task $T_i$.
To ensure the correctness of this method, each predicate must satisfy the following properties:

\begin{itemize}
    
    \item $Prop_1$: each predicate $P_i$ must be computable based on the configuration $C$ observed during the {\tt Look} phase; 
    
    \item$Prop_2$: the predicates must be mutually exclusive, i.e., $P_i \wedge P_j = \texttt{false}$ for all $i \neq j$, ensuring that each robot unambiguously selects a single task;
    
    \item$Prop_3$: for every possible configuration $C$, there must be at least one predicate $P_i$ that is evaluated as true.
    
\end{itemize}

To define each predicate $P_i$, we first identify a set of basic variables that capture relevant properties of the configuration $C$, e.g., metric, numerical, ordinal, or topological features, that can be computed by each robot using only its local observation. Then, such variables are used to assemble a pre-condition $\texttt{pre}_i$ that must be satisfied for task $T_i$ to be applicable. Finally, predicate $P_i$ can be defined as 
$P_i = \texttt{pre}_i \wedge \neg(\texttt{pre}_{i+1} \vee \texttt{pre}_{i+2} \vee \dots \vee \texttt{pre}_q)$.
This definition guarantees that $Prop_2$ is always satisfied and imposes a specific order on the task evaluation. Specifically, predicates are evaluated in reverse order: the robot first checks $P_q = \texttt{pre}_q$, then $P_{q-1} = \texttt{pre}_{q-1} \wedge \neg \texttt{pre}_q$, and so on. If all predicates from $P_q$ down to $P_2$ evaluate to false, then $P_1$ must be true and task $T_1$ is executed.
When a robot performs a generic task $T_i$ in a configuration $C$, the algorithm may lead to a new configuration $C’$ where another task $T_j$ must be performed. In such a case, we say that the algorithm induces a transition from $T_i$ to $T_j$. Collectively, all such transitions form a directed graph called the \emph{transition graph}. The terminal task, which marks the successful resolution of problem $\mathcal{P}$, must correspond to a sink vertex in this graph.
As shown in~\cite{CDN21a}, the correctness of an algorithm designed in this way is guaranteed if the following conditions hold:
\begin{description}
    \item[$H_1$:] the transition graph is correct, i.e., for each task $T_i$, the reachable tasks are exactly those depicted in the transition graph;
    \item[$H_2$:] all the loops in the transition graph, including self-loops not involving sink vertices, must be executed a finite number of times;
    \item[$H_3$:] with respect to the studied problem $\mathcal{P}$, no configuration a-priori proved unsolvable is generated by 
        $\mathcal{A}$.
\end{description}


\section{Mandatory movements and impossibility results}\label{sub:imposs}
%
In this section, we show that in the context of vertex- and edge-transitive graphs there are basic configurations in which some movements are imposed on every gathering algorithm. Moreover, there are configurations in which Gathering cannot be solved.

\begin{lemma}\label{lem:2v} 
Let $G\in \vet$ and $C=(G,\lambda)$ be a configuration with exactly two robots occupying two distinct vertices $u$ and $v$, respectively. In $C$, every gathering algorithm $\A$ must necessarily move the robots so that their distance is reduced.
\end{lemma}
\begin{proof}
Since $G\in \vet$, when a robot in $C$ runs $\A$, it cannot exploit any topological information other than its distance from the other robot. Assume that in $C$ the distance between the two robots, $r$ lying on vertex $u$ and $r'$ lying on another vertex $v$, satisfies $d(u, v)= t$.

If $\A$ makes $r$ move to a neighbor $u' \in N(u)$ such that $d(u', v) \ge t$, the resulting configuration $C'$ still requires the two robots to get closer, eventually, in order to accomplish Gathering. This means that at some point the distance between the robots must decrease until they reach the same vertex where their distance is null. Since in one round the distance can decrease of at most one unit, it follows that at some point $d(r,r')$ is again equal to $t$. Once this happens, the obtained configuration is isomorphic to $C$. Hence, the two robots would start again the same sequence of moves applied from $C$ without ever reaching a common vertex. 


It follows that, in the initial configuration $C$, algorithm $\A$ must make the active robot move from $u$ to a neighbor $u'' \in N(u)$ such that $d(u'', v) < t$.\qed
\end{proof}

\begin{theorem} \label{teo:vt-ungath} 
Let $G\in \vet$, and consider the \rr\ scheduler. The following types of initial configurations are ungatherable:
\begin{itemize}
    \item configurations with exactly 2 neighboring vertices occupied;
    \item configurations where every vertex is occupied. 
\end{itemize}
\end{theorem}
\begin{proof}
Let $C$ be a configuration with exactly 2 neighboring vertices $u$ and $v$ occupied. In such a case, the adversarial strategy may determine the following: (1) in $C$ there are 3 robots in total, 2 located at $u$ and 1 located at $v$; the first active robot is on $u$, the second is the only one at $v$, and the last is the other robot at $u$. Since robots cannot detect multiplicities, according to Lemma~\ref{lem:2v}, every gathering algorithm must necessarily move the active robot to the other occupied vertex $v$. This would result in an infinite execution, as every new configuration remains isomorphic to $C$, preventing the robots from gathering.

Let $C$ be a configuration where every vertex is occupied and let $n$ be the number of vertices in $G$. In such a case, the adversarial strategy may determine the following: 
\begin{itemize}
\item
In $C$, there are $n+1$ robots in total, there is exactly one vertex (say $u$) with 2 robots, and all the other vertices have 1 robot each; 
\item
Since $G\in \vet$, the only possible move would lead the active robot from one occupied vertex to another occupied vertex, without having the possibility for the algorithm to determine a specific robot or a specific destination vertex. The adversary may determine that the first active robot (i.e., at round 1) is on $u$. Moreover, the active robot at round $t>1$ is located at the vertex selected as destination at round $t-1$, and the destination vertex at round $t$ has been never used as destination in the previous $t-1$ rounds. 
\end{itemize}
Also in this case, this adversarial strategy would result in an infinite execution, as every new configuration remains isomorphic to $C$, preventing the robots from gathering.\qed
\end{proof}

The following statement handles configurations defined on classic examples of vertex-transitive graphs.

\begin{theorem} \label{teo:vt-ungath-complete} 
Let $C=(G,\lambda)$ be an initial configuration. Assume the \rr\ scheduler and one of the following cases:
\begin{itemize}
    \item $G$ is a complete graph $K_n$ with at least two occupied vertices;
    \item $G$ is complete bipartite graph $K_{n,n}$ with at least one occupied vertex per side.
\end{itemize} 
Then, $C$ is ungatherable. 
\end{theorem}
\begin{proof}
Assume $G$ is a complete graph $K_n$ with at least two occupied vertices. Due to the symmetry of the graph, only two moves are possible: toward an unoccupied vertex or toward an occupied vertex. In the former case, by repeating the same move and assuming a sufficient number of robots, a configuration where each vertex is occupied will eventually be reached, which is ungatherable by Theorem~\ref{teo:vt-ungath}. Then, a move toward occupied vertices is mandatory.  In this case, assuming a multiplicity at each occupied vertex, the \rr\ scheduler could lead to an infinite loop (i.e., the adversary can choose a sequence of robot activations such that no vertex becomes unoccupied). Hence, $C$ cannot be gathered. 

Assume now that $G$ is a complete bipartite graph $K_{n,n}$ with at least one occupied vertex per side. Again, by symmetry, only two moves are possible: toward an unoccupied vertex or toward an occupied vertex. Assume also multiplicities in each vertex. Moving toward an unoccupied vertex produces a new configuration $C'$ with the same properties as $C$ (and repeating the same move will lead to a configuration where each vertex is occupied, which is ungatherable by Theorem~\ref{teo:vt-ungath}, as above). Moving toward an occupied vertex does not change the configuration in the presence of multiplicities, and repeating the move, the \rr\ scheduler may lead to an infinite loop. Hence, also in this case, $C$ cannot be gathered.\qed
\end{proof}

Note that if $G$ is a complete bipartite graph $K_{n,n}$ with all the occupied vertices on the same side, then trivially the configuration can be gathered. In fact, when the first epoch starts, the first robot moves to the other side; then, all the remaining robots to be activated apply the following strategy: move to the unique occupied vertex on the other side of $G$. Regardless of possible multiplicities, at the end of the first epoch, Gathering is accomplished. The key point in this case is that the analyzed configuration forms what we are going to call a \emph{nice-star} configuration as defined below. 

\begin{definition} Let $C=(G,\lambda)$ be a configuration. We say $C$ forms a \emph{nice-star} configuration if there exists a vertex $v$ in $G$ such that:
\begin{itemize}
    \item $v$ is unoccupied;
    \item every robot in $C$ occupies a vertex belonging to $N(v)$.
\end{itemize}
\end{definition}

The next theorem  shows that in order to solve Gathering from a configuration on a  graph belonging to $\vet$, necessarily a nice-star configuration must be achieved.

\begin{theorem}\label{th:star}
Let $G\in \vet$ and $C=(G,\lambda)$ be an initial configuration with at least 3 vertices occupied. 
If an algorithm $\A$  solves Gathering from $C$ under \rr, then $\A$ must necessarily create an intermediate nice-star configuration $C^*$ during its execution.
\end{theorem}

\begin{proof}
Starting from $C$, let $t$ be the last time instant during the execution of $\A$ where the reached configuration $C’$ has exactly $3$ vertices occupied. Since $C$ has at least 3 occupied vertices, and $\A$ solves Gathering, the time instant $t$ must exist, possibly $t=0$. By hypothesis, at round $t+1$, the obtained configuration $C''$ has exactly 2 occupied vertices. Any subsequent configuration will have exactly 2 occupied vertices until Gathering is accomplished. We now prove that the two vertices occupied in $C''$, say $v_1$ and $v_2$, have the following properties:
\begin{description}
    \item[i)] $v_1$ and $v_2$ are neighbors;
    \item[ii)] all robots in $v_1$ ($v_2$, resp.) are activated by the \rr\ scheduler, initially chosen by the adversary, in a consecutive way, i.e., without interleaving activations of robots in $v_2$ ($v_1$, resp.).
\end{description}

By contradiction, let us assume (i) is false. By the movement dictated by Lemma~\ref{lem:2v}, any robot must move toward the other occupied vertex. In fact, robots cannot detect they are more than two. Hence, if $v_1$ and $v_2$ are not neighbors and the moving robot is part of a multiplicity (which is not recognizable by the robots), then its movement would create a configuration with 3 vertices occupied, against the hypothesis on round $t$.

By contradiction, let us assume (ii) is false. Again by  Lemma~\ref{lem:2v}, it would be impossible to finalize Gathering as robots move from $v_1$ to $v_2$ and from $v_2$ to $v_1$ in an interleaved way, so as to never make one of those vertices unoccupied.



In order to obtain a configuration that respects properties i) and ii), regardless of the sequence dictated by the chosen \rr\ scheduler, we demonstrate that $\A$ must necessarily lead to a nice-star configuration $C^*$. From $C^*$, in fact, it is easy to see that by making  all the robots move toward the center of the star, within one epoch, first a configuration with exactly 3 vertices occupied forming a 3-vertices path is reached, then a configuration $\hat{C}$ with exactly 2 neighboring vertices occupied. Finally, because of Lemma~\ref{lem:2v} and observing that only the robots that did not move so far from $C^*$ have to move, Gathering is accomplished. In other words, $\hat{C}$ respects properties i) and ii).

In particular, in order to impose both properties i) and ii) in $\hat{C}$, regardless of the activation sequence dictated by the \rr\ scheduler chosen initially by the adversary, it is necessary that, without loss of generality, the robots in $v_1$ have reached such a vertex in consecutive rounds, i.e., they were occupying vertices neighboring $v_1$ (including $v_2$). Furthermore, $v_1$ should have been an unoccupied vertex before the arrival of the robots occupying it in $\hat{C}$ as otherwise the robots already in there could not be ensured to be activated in a consecutive way in the subsequent epoch.
It follows that  configuration $C^*$ before such movements was a nice-star configuration, with $v_1$ being the empty center of the star.\qed
\end{proof}






Let $G\in \vet$ and $v\in V(G)$. Consider $P_3^v$ as the set containing all the paths $P_3$ of $G$ having $v$ as central vertex. Depending on $G$, the automorphisms of $G$ may partition $P_3^v$ into one or more equivalence classes. Let $\mathcal{P}_3^v$ denote the set containing all such equivalence classes. As an example, when $G$ is a clique, $\mathcal{P}_3^v$ contains only one element, meaning that all $P_3$ paths centered at $v$ are equivalent. In contrast, 
when $G$ is a square tessellation graph, $\mathcal{P}_3^v$ contains two equivalence classes. 

Any deterministic gathering algorithm $\mathcal{A}$ must select specific elements of $\mathcal{P}_3^v$ -- but not all -- that $\A$ is allowed to generate, as it moves the robots from a nice-star configuration to a gathered one passing through an intermediate $P_3$ configuration. In the following, we denote as $P_{3\mathcal{A}}^v$ the elements of $\mathcal{P}_3^v$ selected by a gathering algorithm $\mathcal{A}$.


\begin{corollary} \label{cor:p3-path}
Let $G\in \vet$ and $C=(G,\lambda)$ be an initial configuration whose occupied vertices form a path in $P_3^v$. 
Then,
\begin{itemize} 
    \item if $|\mathcal{P}_3^v| = 1$,  
    $C$ is ungatherable by any algorithm $\A$;
  \item if $|\mathcal{P}_3^v| > 1$, 
  $C$ is ungatherable by any algorithm $\A$ if the occupied vertices in $C$ form a path in $P_{3\mathcal{A}}^v$.
\end{itemize} 
\end{corollary}

\begin{proof}
According to Theorem~\ref{th:star}, any solving algorithm $\A$, starting from an arbitrary configuration $C'$ with at least three occupied vertices, moves the robots to form a nice-star configuration. Then, it finalizes Gathering by moving the robots to an intermediate configuration $C''$ whose occupied vertices form a $3$-vertex path. 
However, if the occupied vertices of $C'$ form a $3$-vertex path, the algorithm first breaks it to form a nice-star, and then generates $C''$ that is a $3$-vertex path configuration again. However, to finalize Gathering, $C''$ must be distinguishable from $C'$; otherwise the algorithm cannot distinguish between the initial configuration and a generated one. If $C''$ is isomorphic to $C$, the algorithm enters into a loop that prevents Gathering. Hence, if $|\mathcal{P}_3^v| = 1$, the configuration is ungatherable.

If $|\mathcal{P}_3^v| > 1$, the algorithm $\A$ 
generates a configuration $C''$ chosen among those in $P_{3\mathcal{A}}^v$. This implies that, to distinguish between $C''$ and any $C'$ and finalize Gathering, any configuration in $P_{3\mathcal{A}}^v$ has to be excluded from the initial ones.\qed
\end{proof}

\section{Hypercubes}\label{sub:cubes}
In this section, we address Gathering under the \rr\ scheduler when robots move on a $d$-dimensional hypercube. 

\subsection{Notation and terminology} 
The $d$-dimensional hypercube $Q_d$ is the undirected graph with vertex set $V(Q_d) = \{0, 1\}^d$ and two vertices are adjacent if and only if the two tuples differ by exactly one position. Clearly, $Q_d\in \vet$. Hypercube $Q_d$ can also be recursively defined in terms of the Cartesian product of two graphs as follows:

\begin{itemize}
\item
$Q_1 = K_2$,
\item
$Q_d= Q_{d-1} \square\ K_2$, for $d\ge 2$.
\end{itemize}

\begin{figure}[t]
  \centering
  \resizebox{0.8\textwidth}{!}{%
      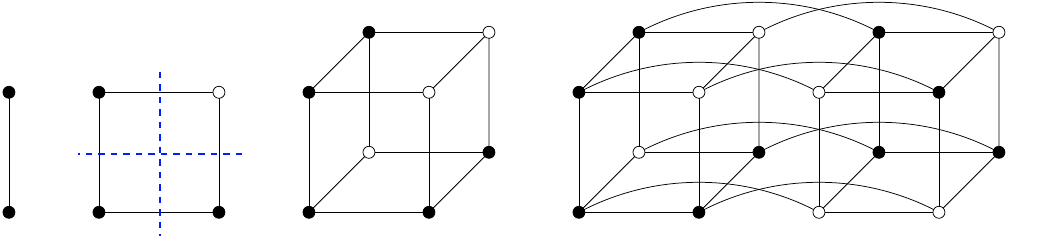
          }
          \caption{A representation  configurations defined on hypercubes $Q_d$ with $d = 1$, $2$, $3$ and $4$. Black vertices represent robots. Dashed blue lines show the two possible partitions of $Q_2$ into hypercubes of dimension one. }
          \label{fig:hypercube}
\end{figure}

According to the definition of the hypercube, the vertices of $Q_d$ can be partitioned into two subsets, each inducing a graph isomorphic to $Q_{d-1}$, by simply applying a cut. Figure~\ref{fig:hypercube} shows examples of configurations defined on $Q_d$, along with two possible cut-sets for $Q_2$. Notice that, for $Q_d$, there are $d$ different cut-sets of $2^{d-1}$ edges each. 

Let $C=(Q_d,\lambda)$ be a configuration. We use $occ(Q')$ to denote the number of occupied vertices of the sub-hypercube $Q'$ (which may coincide with $Q_d$), and $isOcc(v)$ is used as a Boolean value indicating whether a vertex $v$ is occupied or not. 
Given $V'\subset V(Q_d)$, we say that $V'$ is \emph{full} (\emph{empty}, respectively) if each vertex of $V'$ is occupied (unoccupied, respectively). Sometimes, we also use \emph{fully occupied} instead of full. Notation $\mbh(C)$ is used to represent the \emph{minimum bounding hypercube} of $C$, that is, the minimum sub-hypercube of $Q_d$ containing all the occupied vertices. For the sake of notation, we simply use $Q_b$ to denote $\mbh(C)$ (of course, $b\le d$ holds).

\subsection{Algorithm description}
\begin{figure}[htbp]
  \centering
  \resizebox{0.7\textwidth}{!}{%
      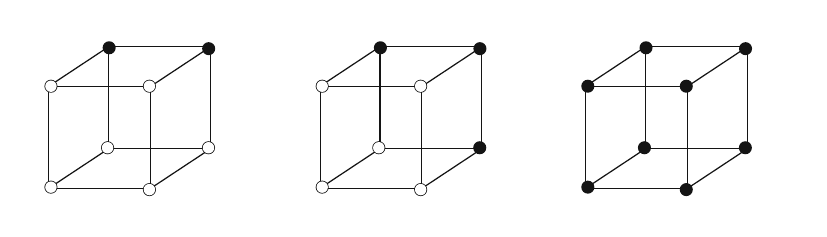
          }
          \caption{Ungatherable initial configurations for $Q_3$. Left: a $P_2$ graph, center: a $P_3$ graph, a fully occupied $Q_3$ graph.}
          \label{fig:ungatherable}
\end{figure}

Let $C=(Q_d,\lambda)$ be any configuration. According to Theorem~\ref{teo:vt-ungath}, $C$ is ungatherable when the occupied vertices induce a $P_2$ or each vertex in $Q_d$ is occupied. Additionally, according to Corollary~\ref{cor:p3-path}, $C$ is ungatherable also when the occupied vertices induce any $P_3$ graph. In fact, it is easy to verify that when the graph is a hypercube, all $P_3$ are equivalent. We denote as $\Uh$ the set containing all these ungatherable configurations (see the examples in Figure~\ref{fig:ungatherable} defined in the context of the hypercube $Q_3$).

In this section, we present $\algohyp$, an algorithm able to gather robots placed on any configuration defined on $Q_d$, $d\ge 3$, and not included in $\Uh$. Notice that for $Q_2$ the hypercube corresponds to a $C_4$ cycle (already studied in~\cite{NP25}) and for $Q_1$ to a path $P_2$ (where {\gath} is either unsolvable or trivial).

The strategy of the proposed algorithm is the following: if the minimum bounding hypercube of $C$ is $Q_b$, with $b>3$, $\algohyp$ moves the robots so that the dimension of the minimum bounding hypercube is reduced to $b-1$. This process is repeated until all the robots are contained within a 3-dimensional hypercube. In this case, the algorithm uses special moves to prevent ungatherable configurations and obtain gathering. 

The reduction of the dimension of the minimum bounding hypercube is obtained as follows: assume $\mbh(C)$ corresponds to a hypercube $Q_b$ with $b>3$, and all robots agree on a specific partition of $Q_b$ into two sub-hypercubes (say $S$ and $D$) of dimension $b-1$. Now, let $r$ be the active robot, $v\in S$ the vertex where $r$ is located, and $v'\in D$ adjacent to $v$. In such a case, the algorithm checks whether a ``direct move'' of $r$ is allowed, that is, if moving $r$ along the ``directed edge'' from $v$, i.e., the edge $(v,v')$, does not generate any ungatherable configuration (or any configuration that conflicts with the strategy, as explained later). If this direct move is allowed, it is performed (and its continuous application will for sure reduce the dimension of the bounding hypercube); otherwise, the analysis of the reason for its inapplicability leads to the use of a special move that, in any case, advances the process. For checking whether a direct move from $S$ to $D$ is allowed, the function $\DMA(S,D)$ (a shorthand for ``Direct Move Allowed'') is defined as follows.


\begin{definition}\label{def:DMA}
Let $S$ and $D$ be two sub-hypercubes of dimension $b-1$ forming a partition of the minimum bounding hypercube $Q_b$. $\DMA(S,D)$ returns false if and only if one of the following cases holds:
\begin{description}
    \item[(a)] $S$ has exactly one occupied vertex and $D$ is full;
    \item[(b)] $S$ has exactly one occupied vertex $v$, and $D$ has exactly one unoccupied vertex $w$ that is adjacent to $v$;
    \item[(c)] $S$ has exactly two adjacent occupied vertices (say $v$ and $v'$), and $D$ has exactly one unoccupied vertex $w$ that is adjacent to $v$.
\end{description}
\end{definition}

\begin{figure}[htbp]
  \centering
  \resizebox{0.9\textwidth}{!}{%
\begingroup%
  \makeatletter%
  \providecommand\color[2][]{%
    \errmessage{(Inkscape) Color is used for the text in Inkscape, but the package 'color.sty' is not loaded}%
    \renewcommand\color[2][]{}%
  }%
  \providecommand\transparent[1]{%
    \errmessage{(Inkscape) Transparency is used (non-zero) for the text in Inkscape, but the package 'transparent.sty' is not loaded}%
    \renewcommand\transparent[1]{}%
  }%
  \providecommand\rotatebox[2]{#2}%
  \newcommand*\fsize{\dimexpr\f@size pt\relax}%
  \newcommand*\lineheight[1]{\fontsize{\fsize}{#1\fsize}\selectfont}%
  \ifx\svgwidth\undefined%
    \setlength{\unitlength}{483.83885193bp}%
    \ifx\svgscale\undefined%
      \relax%
    \else%
      \setlength{\unitlength}{\unitlength * \real{\svgscale}}%
    \fi%
  \else%
    \setlength{\unitlength}{\svgwidth}%
  \fi%
  \global\let\svgwidth\undefined%
  \global\let\svgscale\undefined%
  \makeatother%
  \begin{picture}(1,0.18780487)%
    \lineheight{1}%
    \setlength\tabcolsep{0pt}%
    \put(0,0){\includegraphics[width=\unitlength,page=1]{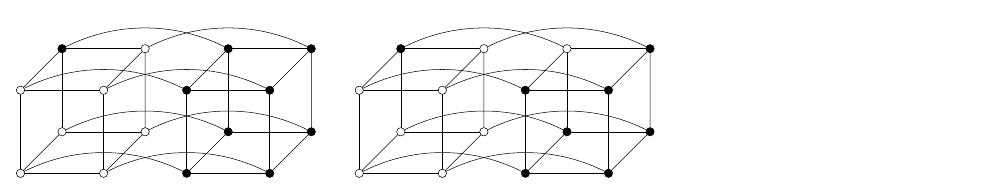}}%
    \put(0.37248123,0.1362063){\color[rgb]{0,0,0}\makebox(0,0)[lt]{\lineheight{1.25}\smash{\begin{tabular}[t]{l}$v$\end{tabular}}}}%
    \put(0.53562284,0.1362063){\color[rgb]{0,0,0}\makebox(0,0)[lt]{\lineheight{1.25}\smash{\begin{tabular}[t]{l}$w$\end{tabular}}}}%
    \put(0,0){\includegraphics[width=\unitlength,page=2]{DMA_ab.pdf}}%
    \put(0.71350386,0.1362063){\color[rgb]{0,0,0}\makebox(0,0)[lt]{\lineheight{1.25}\smash{\begin{tabular}[t]{l}$v$\end{tabular}}}}%
    \put(0.87664552,0.1362063){\color[rgb]{0,0,0}\makebox(0,0)[lt]{\lineheight{1.25}\smash{\begin{tabular}[t]{l}$w$\end{tabular}}}}%
    \put(0.67010098,0.09590362){\color[rgb]{0,0,0}\makebox(0,0)[lt]{\lineheight{1.25}\smash{\begin{tabular}[t]{l}$v'$\end{tabular}}}}%
  \end{picture}%
\endgroup%

          }
          \caption{From left, configurations for which the function $\DMA$ returns false according to conditions (a), (b), (c), respectively.}
          \label{fig:DMA}
\end{figure}

Figure~\ref{fig:DMA} shows configurations illustrating the conditions under which the function $\DMA$ is evaluated as false. We use the notation $\overline{\DMA.(x)}$ to state that the function $\DMA$ returns false by condition $(x)$, $x\in \{a,b,c\}$. When the above function returns true, any active robot $r$ in $S$ can make a direct move to $D$.
%
%
One clear difficulty with this approach is that it is not always possible to uniquely partition $Q_b$ into the sub-hypercubes $S$ and $D$. To overcome this difficulty, the algorithm computes and exploits the following sets:

\begin{definition}
Let $Q_b$ be the minimum bounding hypercube enclosing all robots of a configuration $C$. Then:
\begin{itemize}
    \item $L_0$ contains all the pairs $(S,D)$ representing possible partitions of $Q_b$ into two sub-hypercubes of dimension $b-1$ and such that $D$ has maximum number of occupied vertices;
    \item $L_1$ is the subset of $L_0$ with pairs $(S,D)$ such that $occ(S) < occ(D)$;
    \item $L_2$ is the subset of $L_1$ with pairs $(S,D)$ such that $\DMA(S,D)$ is true; 
    \item $L_3$ is the subset of $L_2$ with pairs $(S,D)$ such that a robot $r$ in $S$ can reach an unoccupied vertex of $D$ with one move.
\end{itemize}
\end{definition}

\begin{algorithm}[t]
\small
\SetKwInput{Proc}{Algorithm}
\Proc{$\algohyp$}
\SetKwInOut{Input}{Input}
\Input{A configuration $C=(Q_d,\lambda)$, $b\ge 3$, such that $C\not\in \Uh$.}
\BlankLine
\BlankLine
    Compute $Q_b = \mbh(C)$ \label{l:computeQk}\;
    
    \lIf{ $b=3$ }{  \fbox{Task $T_1$}}
    \Else(\tcp*[h]{ here, $b>3$ }){
        Compute the sets $L_0$, $L_1$, $L_2$, and $L_3$ \;
        \If{ $|L_1|>0$ }{
            \If{ $|L_2|>0$ }{
                \lIf{ $|L_2|=1$ }{
                    \fbox{Task $T_2$}
                } 
                \Else(\tcp*[h]{here, $|L_2|>1$}){
                    \lIf{ $L_3 > 0$ }{
                        \fbox{Task $T_3$}               
                    } 
                    \lElse(\tcp*[h]{here, $|L_3|=0$}){
                        \fbox{Task $T_4$}
                    }
                }
            }
            \Else(\tcp*[h]{here, $|L_2|=0$}){
                \lIf{ $\overline{\DMA.(a)}$ holds }{
                    \fbox{Task $T_{5.i}$}
                }
                \lIf{ $\overline{\DMA.(b)}$ holds }{
                    \fbox{Task $T_{5.ii}$}
                }
                \lIf{ $\overline{\DMA.(c)}$ holds }{
                    \fbox{Task $T_{5.iii}$}
                }                
            }
        }
        \Else(\tcp*[h]{ here, $L_1=0$ }){
            \If{  
            exists $(S,D)\in L_0$ such that $S$ or $D$ have unoccupied vertices    
            }{
                 \lIf{ exists an edge $(v,w)$ such that $v\in S$ is occupied and $w\in D$ is unoccupied  }{
                    \fbox{Task $T_6$}
                 }
                 \lElse{
                    \fbox{Task $T_7$}
                 }
            }        
            \lElse{
                \fbox{Task $T_8$}
            }  
        }
    }
\caption{ Global structure of $\algohyp$ and its decomposition into tasks. }
\label{alg:algohyp}
\end{algorithm}
  
Algorithm $\algohyp$ solves Gathering by subdividing the problem into ten distinct tasks. These tasks emerge from the global structure of $\algohyp$ as described in the pseudo-code presented in Figure~\ref{alg:algohyp}. The first task (i.e., Task $T_1$, see line 2 of the pseudo-code) handles the case in which the minimum bounding hypercube $Q_b$ of the initial configuration has dimension $b=3$ (hence $\pre_1 \equiv [b=3]$). The other tasks are dedicated to the general case in which $b>3$, and  are defined according to the sets $L_0$--$L_3$. For instance, at line 7, it is possible to observe when Task 2 is activated: its pre-condition corresponds to $\pre_2 \equiv [b >3 \wedge |L_1|>0 \wedge |L_2|=1]$. 

According to the way the pre-conditions of the tasks are defined, and also to the definition of the four sets $L_0$--$L_3$, we have that $\algohyp$ is well defined.
%

We now formalize each of the tasks. For each task, we first recall the pre-condition derived from Figure~\ref{alg:algohyp}, then we describe the rationale of the task, and finally we formalize the move planned for the task. We start from Task $T_2$ and proceed until Task $T_9$. At the end, we give full details about task $T_1$, which is dedicated to gathering robots when they are confined in a minimum bounding hypercube of dimension 3.

\block{Task $T_2$} In this task, there are many pairs $(S,D)$ for which $occ(S)<occ(D)$ but, among them, exactly one pair guarantees that a direct move from $S$ to $D$ is allowed. The precondition associated with this task is as follows:
 $$   \pre_2  \equiv  [ b >3 \wedge |L_1| > 0 \wedge |L_2| = 1]. $$
The corresponding move $m_2$ requires that the active robot $r$ located in $v\in S$ that can perform a direct move (via the direct edge from $v$) moves toward $D$. 

\block{Task $T_3$} In this task, there are several pairs $(S,D)$ having a direct edge toward an empty vertex in $D$. Thus, the algorithm chooses any $(S,D) \in L_3$, and selects any robot $r$ located in  $v\in S$ such that $v$ has an adjacent empty neighbor $v'$ in $D$.
The precondition of this task is the following:
$$   \pre_3   \equiv  [ b >3 \wedge |L_1| > 0 \wedge  |L_2| > 1 \wedge |L_3| > 0]. $$
The move planned for this task is denoted as $m_3$ and it simply requires that $r$ moves to $v'$.

\block{Task $T_4$} In this task the set $L_3$ is empty, and the precondition is as follows:
$$   \pre_4   \equiv  [ b >3 \wedge |L_1| > 0  \wedge |L_2| > 1 \wedge |L_3| = 0]. $$
The algorithm chooses any robot $r$ that resides at a vertex of some $S$ that has empty neighbors within $S$, and then applies move $m_4$: $r$ moves toward $D$.

\block{Tasks $T_{5.i}$, $T_{5.ii}$, and $T_{5.iii}$} 
We now consider the case when $|L_2|=0$. This happens when the function $\DMA$ returns false. Since this depends on three different cases, three different tasks are planned.
In the first of such tasks, function $\DMA$ is false according to condition (a): $S$ has exactly one occupied vertex $v$ and $D$ is full. 
$$   \pre_{5.i}   \equiv  [ b >3 \wedge |L_1| > 0 \wedge |L_2| = 0 \wedge \overline{\DMA.(a)}]. $$
Let $v' \in D$ be the vertex adjacent to $v$, then the planned move $m_{5.i}$ requires that a robot $r$ located on $v'$ moves toward $v$.

In task $T_{5.ii}$, $\DMA$ is false according to condition (b): there exists a unique pair $(S,D)$ such that $S$ has a single occupied vertex $v$ and $D$ has a single empty vertex $w$ adjacent to $v$. If the robot $r$ on $v$ were moved to $w$, $D$ would be filled, thus resulting in an unsolvable configuration. Instead, $r$ is moved to an arbitrary neighbour of $v$ in $S$. This task is activated when the following precondition holds:
$$ \pre_{5.ii}   \equiv  [ b >3 \wedge |L_1| > 0 \wedge |L_2| = 0 \wedge \overline{\DMA.(b)}]$$ 
The corresponding move $m_{5.ii}$ requires that the robot at $v$ moves to an arbitrary neighbour in $S$.

In task $T_{5.iii}$, $\DMA$ is false according to condition (c): $S$ has exactly two adjacent occupied vertices (say $v$ and $v'$), and $D$ has exactly one empty vertex $w$ that is adjacent to $v$.
The precondition for this task is as follows:
$$   \pre_{5.iii}   \equiv  [ b >3 \wedge |L_1| > 0 \wedge |L_2| = 0 \wedge \overline{\DMA.(c)}] $$
The corresponding move $m_{5.iii}$ requires that any robot located in $v$ moves  toward $v'$.
%
%

\begin{figure}[t]
  \centering
  \resizebox{0.75\textwidth}{!}{%
\begingroup%
  \makeatletter%
  \providecommand\color[2][]{%
    \errmessage{(Inkscape) Color is used for the text in Inkscape, but the package 'color.sty' is not loaded}%
    \renewcommand\color[2][]{}%
  }%
  \providecommand\transparent[1]{%
    \errmessage{(Inkscape) Transparency is used (non-zero) for the text in Inkscape, but the package 'transparent.sty' is not loaded}%
    \renewcommand\transparent[1]{}%
  }%
  \providecommand\rotatebox[2]{#2}%
  \newcommand*\fsize{\dimexpr\f@size pt\relax}%
  \newcommand*\lineheight[1]{\fontsize{\fsize}{#1\fsize}\selectfont}%
  \ifx\svgwidth\undefined%
    \setlength{\unitlength}{414.21327209bp}%
    \ifx\svgscale\undefined%
      \relax%
    \else%
      \setlength{\unitlength}{\unitlength * \real{\svgscale}}%
    \fi%
  \else%
    \setlength{\unitlength}{\svgwidth}%
  \fi%
  \global\let\svgwidth\undefined%
  \global\let\svgscale\undefined%
  \makeatother%
  \begin{picture}(1,0.29629628)%
    \lineheight{1}%
    \setlength\tabcolsep{0pt}%
    \put(0,0){\includegraphics[width=\unitlength,page=1]{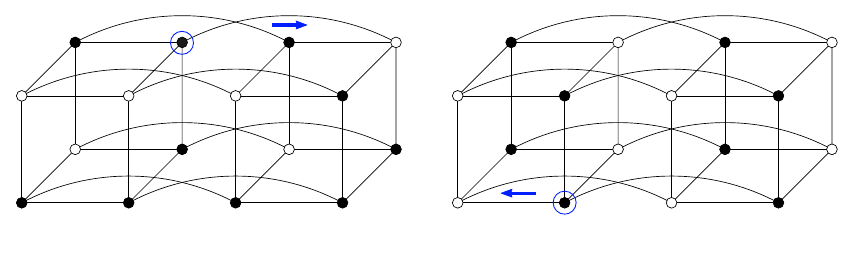}}%
    \put(0.06554444,0.01933558){\color[rgb]{0,0,0}\makebox(0,0)[lt]{\lineheight{1.25}\smash{\begin{tabular}[t]{l}$S$\end{tabular}}}}%
    \put(0.31179438,0.01933558){\color[rgb]{0,0,0}\makebox(0,0)[lt]{\lineheight{1.25}\smash{\begin{tabular}[t]{l}$D$\end{tabular}}}}%
    \put(0.57615097,0.01933558){\color[rgb]{0,0,0}\makebox(0,0)[lt]{\lineheight{1.25}\smash{\begin{tabular}[t]{l}$S$\end{tabular}}}}%
    \put(0.82240092,0.01933558){\color[rgb]{0,0,0}\makebox(0,0)[lt]{\lineheight{1.25}\smash{\begin{tabular}[t]{l}$D$\end{tabular}}}}%
  \end{picture}%
\endgroup%

          }
          \caption{Example configurations. Left: a configuration in $T_{6}$, right: a configuration in $T_{7}$. The moving robot is highlighted with a circle. The arrow shows the movement direction. }
          \label{fig:task9}
\end{figure}

\block {Tasks $T_6$ and $T_7$}
These tasks address the cases in which $|L_1|=0$ (hence, $occ(S)=occ(D)$ holds for each possible pair $S,D)$). If there exists a pair $(S,D)\in L_0$ such that $S$ or $D$ has unoccupied vertices, then Task $T_6$ starts; otherwise, $T_7$ is responsible for managing the configuration. Given 
$\pre' \equiv [\exists (S,D) \in L_0: occ(S\cup D) < |V(Q_b)| ]$ and 
$\pre'' \equiv [  \exists (v,w)\in (S\times D): isOcc(v) \wedge \neg{isOcc(w)}]$, the pre-condition of $T_6$ can be formalized as follows. 
%
%
$$   \pre_{6}   \equiv  
     [ b >3 \wedge |L_1| = 0 
       \wedge \pre' 
       \wedge \pre''].$$
The corresponding move $m_{6}$ simply requires moving a robot from $v$ to $w$ when $(v,w)\in (S\times D)$ is the pair involved in $\pre''$ (see Figure~\ref{fig:task9}, left). The pre-condition of $T_7$ becomes:
$$   \pre_{7}   \equiv  
     [ b >3 \wedge |L_1| = 0 
       \wedge \pre' 
       \wedge \neg\pre''].$$
It can be easily observed that this condition is equivalent to saying that for each pair $(S,D)\in L_0$ and for each pair of vertices $(v,w)\in (S\times D)$, both $v$ and $w$ are either occupied or unoccupied. The associated move $m_{7}$ is designed to break this ``symmetry''. In particular, any active robot in $S$ moves toward an adjacent vertex in $S$ which is unoccupied (the presence of such an unoccupied vertex is guaranteed by $\pre'$). As an example, see Figure~\ref{fig:task9}, right.

\block{Task $T_8$} In this task, both $S$ and $D$ have all vertices occupied for each pair $(S,D) \in L_0$. Formally:
$$   \pre_{8}  \equiv [b >3 \wedge |L_1| = 0 
               \wedge  \forall (S,D)\in L_0,~ occ(S\cup D) = |V(Q_b)| ].$$
This is equivalent to saying that each vertex in the whole minimum bounding hypercube $Q_b$ is occupied. In such a situation, there exists one or more hypercubes $Q'_b$ (hence, having the same dimension as $Q_b$) directly connected to $Q_b$. The move is designed to enlarge the minimum bounding hypercube by moving one robot outside $Q_b$.
Formally, move $m_{8}$ simply requires that 
the active robot that has an unoccupied neighbor moves toward that vertex.

\block{Task $T_1$} 
The precondition of this task is 
$ \pre_{1} \equiv [b\leq 3]$.
This implies that the minimum bounding cube $Q_b$ is such that $b=3$. All handled configurations are reported in Figure~\ref{Fig:algo-Q3}, together with the move planned by $\algohyp$ in each specific case. In particular, the figure displays a transition graph where vertices represent configurations and edges show possible transitions between them. It includes all distinct configurations of robots on $Q_3$ up to isomorphisms, organized according to the number of robots.
Each node in the transition graph, shows which robot moves and in which direction. Solid arrows represent transitions that occur when a robot moves from a vertex occupied solely by that robot. In contrast, dashed arrows illustrate scenarios where the vertex was initially occupied by multiple robots, and after the move, the configuration allows for one additional occupied vertex. 
If a robot moves toward a vertex that is already occupied, it generates a self-loop in the transition graph. However, such self-loops are not shown in the figure to enhance readability.


%
\subsection{Formalization and Correctness} 
According to the methodology recalled in Section~\ref{sec:prob}, 
we know that 
$\algohyp$ is well defined and we prove its correctness by showing that the three properties $H_1$, $H_2$, $H_3$ hold. We prove this by providing a specific lemma for each task. Table~\ref{tab:tasks} reports all the transitions among tasks. Concerning $H_2$, each lemma will analyze self-loops only (leaving the analysis of other cycles contained in Table~\ref{tab:tasks} to the final correctness theorem). Concerning $H_3$, notice that the ungatherable configurations in $\Uh$ corresponding to $P_2$ and $P_3$ could be formed only during the execution of $T_1$, hence in task $T_2,\ldots,T_9$, property $H_3$ will be checked only against the ungatherable configuration corresponding to the fully occupied $Q_b$. 

\begin{table}[h]
    \centering
    \normalsize
    \begin{tabular}{c|l}
    \hline
$T_1$  & {\gath} \\\hline
$T_2$  & $T_2$, $T_{5.i}$, $T_{5.ii}$, $T_{5.iii}$, $T_*$\\
$T_3$  & $T_2$, $T_3$, $T_4$, $T_{5.i}$, $T_{5.ii}$, $T_{5.iii}$, $T_*$\\
$T_4$  &  $T_2$, $T_4$, $T_{5.i}$, $T_{5.ii}$, $T_{5.iii}$, $T_*$\\\hline
$T_{5.i}$  & $T_{5.i}$, $T_{5.ii}$\\
$T_{5.ii}$  & $T_2$, $T_{5.iii}$\\
$T_{5.iii}$  & $T_{2}$, $T_{5.iii}$\\\hline
$T_6$  & $T_2$, $T_3$, $T_4$, $T_{5.i}$, $T_{5.ii}$, $T_{5.iii}$\\
$T_7$  &  $T_2$, $T_3$, $T_4$, $T_{5.i}$,  $T_{5.ii}$, $T_{5.iii}$, $T_6$\\
$T_8$  & $T_{5.i}$, $T_{5.ii}$\\\hline
    \end{tabular}
\vspace{0.7em}
\caption{
A tabular representation of the transition graph of $\algohyp$.
The first column reports the task's name, and the second column reports the transitions. Task $T_*$ stands for any task and represents the situation in which the dimension of the minimum bounding cube decreases by one.}
\label{tab:tasks}
\end{table}

 \begin{lemma}\label{lem:corr-2}
 Let $C$ be a configuration where $T_{2}$ is applied. Starting from $C$, $\algohyp$ eventually either reduces the size of $\mbh(C)$ or  leads to a configuration where one among tasks $T_{2}$, $T_{5.i}$, $T_{5.ii}$, $T_{5.iii}$ is applied.
\end{lemma}
\begin{proof}
During this task, $pre_{2}$ holds, and an active robot belonging to $S$ moves toward $D$ via a direct edge.  

\begin{description}
\item[$H_1$:] 

After the move, if the size of $\mbh(C)$ is not reduced, then the algorithm re-enters in $T_{2}$ if $\DMA$ holds and $r$ belonged to a multiplicity; the algorithm enters $T_{5.i}$, $T_{5.ii}$, or $T_{5.iii}$ when $\overline{\DMA.(a)}$, $\overline{\DMA.(b)}$, or $\overline{\DMA.(c)}$ holds, respectively. 
Note that, there are no transitions from $T_{2}$ to $T_{3}$ or $T_4$; that is, $|L_2|$ does not increase. Suppose that $|L_2| > 1$ holds after a move. In this case, there must have been a pair $T' = (S', D')$ in addition to the pair $T = (S, D)$ used in the move. Both pairs had the same occupancy in sets $D$ and $D'$, but while pair $T$ allowed the move, pair $T'$ did not. Therefore, $D'$ must have had either no empty vertices or exactly one empty vertex 
in the whole minimum bounding hypercube. Since $D$ and $D'$ have the same number of occupied vertices, this scenario cannot occur for $b > 3$.

\item[$H_2$:] After a move, either $|L_2| = 1$ or $|L_2| = 0$. When $|L_2| = 1$, a self-loop occurs; however, this can happen no more than the number of robots in $S$, thereby making it finite.

\item[$H_3$:] When this task is activated, $occ(S)<occ(D)$ and function $\DMA$ is true. Therefore, after the move, the obtained configuration cannot be the fully occupied $Q_b$.
\end{description}\qed
\end{proof}

\begin{lemma}\label{lem:corr-}
Let $C$ be a configuration where $T_{3}$ is applied. Starting from $C$, $\algohyp$ eventually either reduces the size of $\mbh(C)$ or leads to a configuration where one among tasks $T_{2}$, $T_{3}$, $T_{4}$, $T_{5.i}$, $T_{5.ii}$, and $T_{5.iii}$ is applied. 
\end{lemma}
\begin{proof}
During this task, the algorithm chooses any $(S,D) \in L_3$, and selects any robot $r \in S$ such that $r$ has an adjacent empty neighbour $v$ in $D$.

\begin{description}
 \item[$H_1$:] 
After this move,  if the size of $\mbh(C)$ is not reduced, the algorithm enters 
\begin{itemize}
\item task $T_{2}$ if $\DMA$ holds and  $|L2|=1$, 
\item task $T_{3}$ if $\DMA$ holds, $|L2|>1$, and $|L3| > 0$,
\item task $T_{4}$ if $\DMA$ holds, $|L2|>1$, and  $|L3| = 0$,
\item 
task $T_{5.i}$ ($T_{5.ii}$, $T_{5.iii}$, respectively) when $\overline{\DMA.(a)}$ ($\overline{\DMA.(b)}$, $\overline{\DMA.(c)}$, respectively) holds. 
\end{itemize}
%
\item[$H_2$:] Since $S$ has fewer occupied vertices than $D$, when a robot $r$ moves from $S$ to $D$, the number of occupied vertices of $D$ increases. Even if later, another partition $(S', D')$ may be chosen, the new move would move the robot into $D \cap D'$. This implies that in a finite number of moves, one reaches a partition in which either the move is not allowed or $S$ becomes empty (and thus $b$ decreases).

\item[$H_3$:] 
When this task is activated, $occ(S)<occ(D)$ and function $\DMA$ is true. Therefore, after the move, the obtained configuration cannot be the fully occupied $Q_b$.
\end{description}\qed
\end{proof}

\newpage
\begin{figure}[ht]
      \centering
  \resizebox{0.9\textwidth}{!}{%
      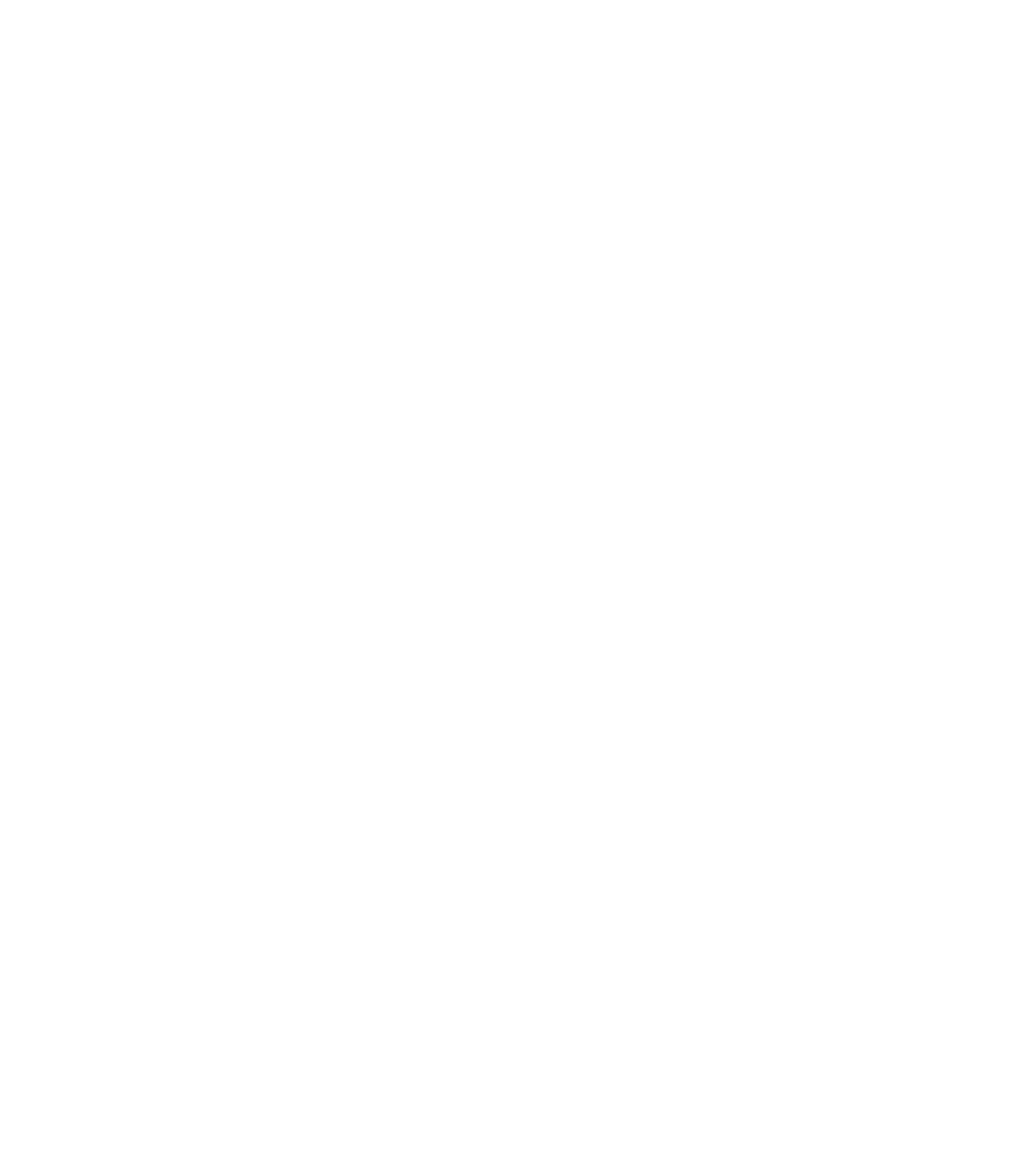
  }
\caption{Configurations and moves planned by $\algohyp$ for Task $T_1$. The graph shows the transitions between configurations. Blue arrows show robot movements. Note that a self-loop exists in each node where a robot moves toward an occupied vertex, but they are omitted for presentation. Dashed arrows mean that the transitions are generated by  robots moving from vertices occupied by multiplicities.
}
\label{Fig:algo-Q3}
\end{figure}

\begin{lemma}\label{lem:corr-4}
Let $C$ be a configuration where $T_{4}$ is applied. Starting from $C$, $\algohyp$ eventually either reduces the size of $\mbh(C)$ or leads to a configuration where one among tasks $T_{2}$, $T_{4}$, $T_{5.i}$, $T_{5.ii}$, and $T_{5.iii}$ is applied. 
\end{lemma}
\begin{proof}
During this task $\pre_{4}$ holds, the algorithm chooses any robot $r$ that resides at a vertex of some $S$ with empty neighbors within $S$ and then moves $r$ toward $D$ via a direct move.
\begin{description}
\item[$H_1$:] 
After this move,  if the size of $\mbh(C)$ is not reduced, the algorithm enters
\begin{itemize}
\item task $T_{2}$ if $\DMA$ holds and $|L2|=1$, 
\item task $T_{4}$ if $\DMA$ holds, and $|L2|>1$, and  $|L3| = 0$.
\item $T_{5.i}$ ($T_{5.ii}$, 
 $T_{5.iii}$, respectively) when $\overline{\DMA.(a)}$ ($\overline{\DMA.(b)}$, $\overline{\DMA.(c)}$, respectively) holds. 
\end{itemize}
%

\item[$H_2$:] In Task $T_{4}$, with the initial partition $(S,D)$, with $occ(S)<occ(D)$, after the first move from $S$ to $D$ — even in the worst-case scenario where the robot moves from a vertex with multiplicity — there can be no further moves along the edges connecting $S$ to $D$ in the opposite direction. Note that these edges are the only ones that could allow the first moved robot to return to its initial position and potentially create an infinite loop.
\item[$H_3$:] When this task is activated, $occ(S)<occ(D)$ and function $\DMA$ is true. Therefore, after the move, the obtained configuration cannot be the fully occupied $Q_b$.
\end{description}\qed
\end{proof}

\begin{lemma}\label{lem:corr-5.i}
Let $C$ be a configuration where $T_{5.i}$ is applied. Starting from $C$, $\algohyp$ eventually leads to a configuration where either Task $T_{5.i}$ or $T_{5.ii}$ is applied. 
\end{lemma}
\begin{proof}
In this task, it is selected a pair $(S,D)$ such that $S$ has exactly one occupied vertex $v$ and $D$ is full. Let $v' \in D$ be the vertex adjacent to $v$. The algorithm selects the robot $r$ on $v'$ and moves it toward $v$. The dimension of the minimum bounding hypercube does not decrease.
\begin{description}
\item[$H_1$:] 
Since $occ(S) =1$ and $D$ is fully occupied, the algorithm must repeat the same task until the vertex from which the move is initiated becomes empty. Once this vertex is unoccupied, the current task ends, and the task switches to $T_{5.ii}$. Hence, there are transitions only to $T_{5.i}$ or $T_{5.ii}$.
\item[$H_2$:] In Task $T_{5.i}$, since $D$ is fully occupied and $occ(S) = 1$, the algorithm remains in the same task until the vertex from which the move is executed becomes empty. As soon as this occurs, the task terminates.
\item[$H_3$:] In this task, function $\DMA$ is false by condition (a), therefore, the move is designed to empty one vertex in $D$.
Hence, the obtained configuration cannot be the fully occupied $Q_b$.
\end{description}\qed
\end{proof}


\begin{lemma}\label{lem:corr-5.ii}
Let $C$ be a configuration where $T_{5.ii}$ is applied. Starting from $C$, $\algohyp$ eventually leads to a configuration where either Task $T_{5.iii}$ or $T_{2}$ is applied. 
\end{lemma}
\begin{proof}
In this task, a pair $(S,D)$ is selected such that $S$ has a single occupied vertex $v$ and $D$ has a single empty vertex $w$ adjacent to $v$. The robot $r$ on $v$ is moved to an arbitrary neighbor of $v$ in $S$. The dimension of the minimum bounding hypercube does not decrease.
\begin{description}
\item[$H_1$:] 
From this task, there are only two possible transitions according to the number of robots located on the vertex $v$. 
If there is a multiplicity on $v$, then there will be two adjacent occupied vertices in $S$ after the move. Hence, the algorithm applies $T_{5.iii}$. Conversely, if $r$ was the only robot on $v$, then $\DMA$ becomes true, there is one allowed move, $|L_2|=1$, $|L_1|>0$ as $occ(S)<occ(D)$, and the algorithm applies $T_{2}$. 
\item[$H_2$:] 
As soon as one robot moves, the task terminates.
\item[$H_3$:] In this task, the function $\DMA$ is false by condition (b), and a robot is moved inside $S$, therefore, after the move, 
the configuration cannot be the fully occupied $Q_b$.
\end{description}\qed
\end{proof}

\begin{lemma}\label{lem:corr-5.iii}
 Let $C$ be a configuration where $T_{5.iii}$ is applied. Starting from $C$, $\algohyp$ eventually leads to a configuration where either Task $T_{5.iii}$ or $T_2$ is applied. 
\end{lemma}
 \begin{proof}
In this task, $S$ has exactly two adjacent occupied vertices (say $v$ and $v'$), and $D$ has exactly one empty vertex $w$ that is adjacent to $v$. This task continues $T_{5.ii}$ for robots moving from a multiplicity. Any robot in $v$ is moved toward $v'$ until $v$ is empty. The dimension of the minimum bounding hypercube does not decrease.

\begin{description}
\item[$H_1$:] 
If $v$ contains a multiplicity in $C$, then after the move Task $T_{5.iii}$ is still applied. The resulting configuration is essentially the same as $C$ but with one robot removed from $v$ and one added to $v'$. 
Conversely, if $r$ was the only robot on $v$, then $\DMA$ becomes true, there is one allowed move, $|L_2|=1$, $|L_1|>0$ as $occ(S)<occ(D)$, and the algorithm applies $T_{2}$. 
\item[$H_2$:] 
Concerning the self-loop, the algorithm remains in the same task until the vertex from which the move is executed becomes empty. As soon as this occurs, the task terminates.
\item[$H_3$:] in this task, 
$D$ keeps at least one empty vertex, 
hence the created configuration cannot be the fully occupied $Q_b$.
\end{description}\qed
\end{proof}

\begin{lemma}\label{lem:corr-6}
 Let $C$ be a configuration where $T_{6}$ is applied. Starting from $C$, $\algohyp$ eventually leads to a configuration where one among tasks  $T_2$, $T_3$, $T_4$, $T_{5.i}$, $T_{5.ii}$, $T_{5.iii}$ is applied. 
\end{lemma}
\begin{proof}
In this task, $occ(S)=occ(D)$ and the configurations of robots on $S$ and $D$ are not symmetric, therefore exists $v \in S$ that has an unoccupied neighbor in $D$. Any robot in $S$ having an unoccupied neighbor in $D$ moves toward that vertex. The dimension of the minimum bounding hypercube does not decrease.
\begin{description}

\item[$H_1$:] 
After the move, as $occ(S)\neq occ(D)$, the algorithm can apply any task among $T_2$, $T_3$, $T_4$, $T_{5.i}$, $T_{5.ii}$, or $T_{5.iii}$, but not tasks $T_6$, $T_7$, $T_8$, $T_9$.
\item[$H_2$:] This task has no self-loops, and in one robot movement, the configuration is changed so that the precondition of $T_{6}$ is no longer valid.
\item[$H_3$:] 
Since $|L_1|=0$ and $b>3$ hold when the task is applied, the performed move cannot create the fully occupied $Q_b$.
\end{description}\qed
\end{proof}

\begin{lemma}\label{lem:corr-7} 
 Let $C$ be a configuration where $T_{7}$ is applied. Starting from $C$, $\algohyp$ eventually leads to a configuration where one among tasks  $T_2$, $T_3$, $T_4$, $T_{5.i}$, $T_{5.ii}$, $T_{5.iii}$ or $T_6$ is applied. 
\end{lemma}
 \begin{proof}
In this task, $occ(S)=occ(D)$ and the configurations of robots in $S$ are symmetrical to the robots in $D$. Then any robot in $S$ moves toward an unoccupied vertex in $S$. The dimension of the minimum bounding hypercube does not decrease.
\begin{description}
\item[$H_1$:] 
After the move, it can be easily observed that the algorithm may apply any of the tasks $T_2$, $T_3$, $T_4$, $T_{5.i}$, $T_{5.ii}$, $T_{5.iii}$ or even $T_{6}$ if the robot selected for the movement was part of a multiplicity. 
\item[$H_2$:]This task has no self-loops, and, in one move, the precondition of $T_7$ is no longer valid. 
\item[$H_3$:] After the move, either the number of occupied vertices in $S$ remains the same or increases by one. Therefore, 
the move cannot create the fully occupied $Q_b$.
\end{description}\qed
\end{proof}

\begin{lemma}\label{lem:corr-8}
Let $C$ be a configuration where $T_{8}$ is applied. Starting from $C$, $\algohyp$  leads to a configuration where either $T_{5.i}$ or $T_{5.ii}$ is applied. 
\end{lemma}  
 \begin{proof}
In this task, for all $(S,D) \in L_0$, both $S$ and $D$ have all vertices occupied and their union forms the hypercube $Q_b$. The move is designed to enlarge $b$ by moving one robot outside $Q_b$. The dimension of the minimum bounding hypercube will increase.
\begin{description}
\item[$H_1$:] 
After the move, it can be easily observed that
the algorithm either applies task $T_{5.i}$ (if the moving robot was part of a multiplicity) or applies task $T_{5.ii}$ (if $r$ was the only robot on $v$). 
\item[$H_2$:] This task has no self-loops. After one robot moves, the precondition of $T_8$ is no longer valid. 
\item[$H_3$:] 
According to the performed move, it is clear that the fully occupied $Q_b$ cannot be generated.
\end{description}\qed
\end{proof}

\begin{lemma}\label{lem:corr-1}
 Let $C$ be a configuration where $T_1$ is applied.  Starting from $C$, $\algohyp$ eventually leads to solve {\gath}.
 \end{lemma}
 \begin{proof}
In this task, the minimum bounding cube $Q_b$ is such that $b=3$. All the possible managed configurations are reported in Figure~\ref{Fig:algo-Q3}, along with the move planned by $\algohyp$ in each specific case. It includes all distinct configurations of robots on $Q_3$ up to isomorphisms, organized according to the number of occupied vertices. Moreover, the figure displays a transition graph where nodes represent configurations and edges show possible transitions between them. 

The transitions show that there are no cycles apart from self-loops (a self-loop exists in each node where a robot moves toward an occupied vertex). However, the number of such occurrences is bounded by the number of robots at the vertex, thereby ensuring finiteness.

According to the planned moves, the number of occupied vertices always decreases except when the robot moves from vertices occupied by a multiplicity 
(cf. transitions depicted by dashed arrows). The algorithm then gathers the configuration into a single vertex. 

It is worth remarking that, although configurations at nodes labeled 2.1 and 3.3 are elements of $\Uh$, they are generated by $\algohyp$ starting from the configuration labeled 2.2 (which configuration between 2.1 and 3.3 is generated depends on possible multiplicities). It is clear that, starting from configuration labeled 2.2 and within an epoch of robots' activations, the configuration labeled 1.1 will surely be generated. This proves that Task $T_1$ always finalizes the requested gathering.\qed
\end{proof}

The final result can be summarized by the following statement.
 \begin{theorem}\label{teo:hyper}
 Let $C=(Q_d,\lambda)$ be an initial configuration where $Q_d$ is a hypercube graph. 
 The {\gath} problem can be optimally solved in $C$ under the \rr\ scheduler if and only if $C\not\in \Uh$.
 \end{theorem} 
\begin{proof}
According to Theorem~\ref{teo:vt-ungath} and Corollary~\ref{cor:p3-path}, we know that $C$ is ungatherable when $C\in \Uh$. 

In what follows, we assume that $C\not\in \Uh$. 
Lemmata~\ref{lem:corr-2}-\ref{lem:corr-1} ensure that properties $H_1$, $H_2$, $H_3$, hold for all tasks $T_{1},\ldots, T_{8}$. As a consequence, the algorithm never brings $C$ into an unsolvable configuration (except those correctly handled by $T_1$, see the proof of Lemma~\ref{lem:corr-1}). All the transitions are those reported in Table~\ref{tab:tasks}, and generated configurations can only be managed by the same task a finite number of times. Concerning cycles formed by different tasks,  Table~\ref{tab:tasks} indicates that the following are the only cycles generated by $\algohyp$:
\begin{itemize}
\item 
$T_{2} \rightarrow T_{5.i} \rightarrow T_{5.ii} \rightarrow T_{2}$;
\item
$T_{2} \rightarrow T_{5.ii} \rightarrow T_{2}$;
\item 
$T_{2} \rightarrow T_{5.ii} \rightarrow T_{5.iii} \rightarrow T_{2}$;
\item 
$T_{2} \rightarrow T_{5.iii} \rightarrow T_{2}$;
\end{itemize}

To prove that the first two cycles do not generate infinitely many activations, we show that the transition $T_{5.ii} \rightarrow T_{2}$ is traversed only a finite number of times. Indeed, starting from $T_{5.ii}$, if the algorithm re-applies $T_{2}$, then there was a unique robot on the single occupied vertex in $S$. The subsequent move, dictated by $T_{2}$, is executed with respect to the same partition $(S,D)$ and will empty $S$. This terminates the execution relative to the current value of $b$. Concerning the last two cycles, if the algorithm restarts $T_2$ after processing $T_{5.iii}$, then there is only one vertex occupied in $S$. Consequently, in $T_2$ all the robots from the only remaining occupied vertex $v$ in $S$ move to the adjacent vertex in $D$ until decreasing the dimension of $Q_b$. This clearly repeats a number of times equal to the number of robots occupying $v$. Afterwards, the execution corresponding to the current value of $b$ terminates.

According to the transitions reported in Table~\ref{tab:tasks} and to the prove that there are no cycles among transitions that generate infinitely many executions, we get that in a finite number of epochs $\algohyp$ creates a configuration managed by $T_1$. As proved by Lemma~\ref{lem:corr-1}, any configuration in $T_1$ (excluding those in $\Uh$ not generated by the algorithm) is transformed by $\algohyp$ into a final configuration where {\gath} is solved. 


Let us now analyze the time complexity of $\algohyp$. 
The algorithm progressively moves robots from the sub-hypercube $S$ toward $D$, thus either reducing the dimension $b$ of the minimum bounding cube or entering a configuration in which the direct move is not allowed, that is tasks $T_{5.i}$, $T_{5.ii}$, $T_{5.iii}$.
When the direct move is not allowed, each task among $T_{5.1}$, $T_{5.ii}$ or $T_{5.iii}$ requires at most one epoch to change task. Indeed, within one epoch an entire multiplicity is moved to the destination. Hence, in at most three epochs the value of $b$ is reduced by one. 
Note that each of the tasks $T_2$, $T_3$, $T_4$, $T_6$, $T_7$, $T_8$ cannot require more than $O(1)$ epochs before either reducing $b$ or entering one of the tasks in $T_{5.i}$, $T_{5.ii}$, or $T_{5.iii}$. Self-loops are always resolved within one epoch as all the robots composing a multiplicity that are allowed to move, perform their movements within one epoch. Furthermore, Task $T_1$ requires at most 9 transitions (that's the measure of the longest path from the transition graph of Figure~\ref{Fig:algo-Q3}) to finalize Gathering. Each transition requires at most one epoch. 

Therefore, the overall number of epochs required by the algorithm to gather $C$ is $O(diam(Q_d))$, that is $O(1)$ for each dimension from $d$ down to $3$. In fact, according to Lemma~\ref{lem:lb} and by remembering that any hypercube $Q_d$ has $diam(Q_d)=d$, the optimality of the proposed algorithm follows. \qed
\end{proof}



\section{Square Tesselation graphs}\label{sec:grids-tmp}

In this section, we address Gathering under the RR scheduler when robots move on square tessellation graphs.

\subsection{Notation and terminology} 
\begin{figure}[t]
\centering
\includegraphics[scale=0.65]{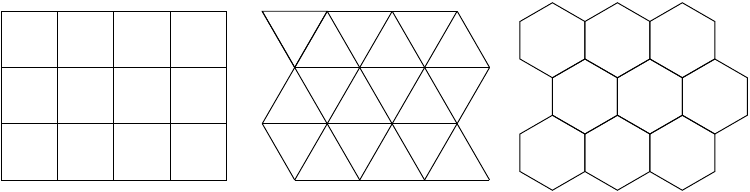}
\caption{Parts of regular plane tessellations.}
\label{fig:tessellation}
\end{figure}

We consider here infinite graphs generated by a \emph{plane tessellation}. A tessellation is a tiling of a plane with polygons without overlapping. A \emph{regular tessellation} is a tessellation which is formed by just one kind of regular polygon of side length $1$ and in which the corners of the polygons are identically arranged. According to~\cite{GS87}, there are only three regular tessellations, and they are generated by squares, equilateral triangles or regular hexagons (see Figure~\ref{fig:tessellation}).

An infinite lattice of a regular tessellation is a lattice formed by taking the vertices of the regular polygons in the tessellation as the points of the lattice. A graph $G$ is induced by the point set $S$ if the vertices of $G$ are the points in $S$ and its edges connect vertices that are distance $1$ apart. A \emph{tessellation graph} of a regular tessellation is the infinite graph embedded into the Euclidean plane 
induced by the infinite lattice formed by that tessellation~\cite{Ionascu12}. We denote by $G_S$ ($G_T$ and $G_H$, resp.) the tessellation graphs induced by the regular tessellations generated by squares (equilateral triangles and regular hexagons, resp.). In this section we consider any configuration $C=(G_S,\lambda)$ where robots are confined within a finite portion of the grid. 

\begin{figure}[t]
  \centering
  \resizebox{0.7\textwidth}{!}{%
      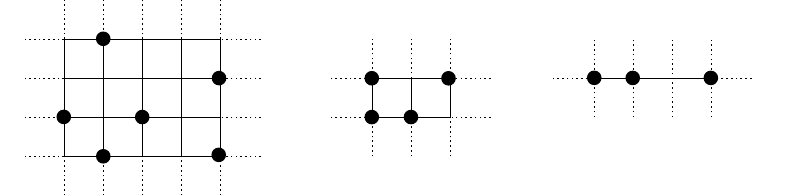
          }
         \caption{From left to right, configurations of type $\ctype{4}{5}$, $\ctype{2}{3}$, and $\ctype{1}{4}$. The first configuration has just one occupied corner. 
         }
\label{fig:conf_4x5}
\end{figure}

Given a configuration $C=(G_S,\lambda)$,
the \emph{minimum bounding rectangle} $\mbr(C)$ is 
the smallest rectangle enclosing all the occupied vertices of $G_S$. We assume $\mbr(C)$ finite. Notice that $\mbr(C)$ is unique and can be easily detected by all robots. We say ``corners of $C$'' to refer to the four corners of $\mbr(C)$. Moreover, we say that $C$ is a $\ctype{m}{n}$-configuration if $\mbr(C)$ encloses $m$ rows and $n$ columns (see Figure~\ref{fig:conf_4x5}). Consequently, in this section, $m$ and $n$ will not denote the number of edges and vertices of a graph, as usual, since they are both infinite, but always the rows and columns of a minimum bounding rectangle.

\subsection{On the ungatherable configurations}
We provide an algorithm $\algosquare$ (Algorithm for Square Tessellation graphs) that is capable of solving the {\gath} problem from each initial $\ctype{m}{n}$-configuration $C=(G_S,\lambda)$, with both $m$ and $n$ bounded, except those that are proved to be ungatherable. 
Concerning the latter, it is worth observing that, according to Theorem~\ref{teo:vt-ungath}, we have to exclude from the possible initial configurations of $\algosquare$ those whose occupied vertices induce a path $P_2$. Notice that there is just one configuration of such a type, and it corresponds to the $\ctype{1}{2}$-configuration. The same theorem includes among the ungatherable configurations the case in which each vertex is occupied, but such a situation does not occur here because we assume $\mbr(C)$ to be finite.
According to Corollary~\ref{cor:p3-path}, we must analyze the configurations with exactly 3 vertices occupied that induce a path $P_3$. These cases correspond to the $\ctype{2}{2}$-configuration with three occupied vertices, and the $\ctype{1}{3}$-configuration with three occupied vertices. Since they can be topologically distinguished, $\algosquare$ must be designed by selecting one of such 3-vertex paths as its ``associated $P_3$ path'' (cf. the discussion just before  Corollary~\ref{cor:p3-path}). Indeed, $\algosquare$ is designed so that its associated path corresponds to the  $P_3$ path in the $\ctype{2}{2}$-configuration with exactly three occupied vertices. 
Summarizing, according to Theorem~\ref{teo:vt-ungath}, Corollary~\ref{cor:p3-path}, and the $P_3$ path associated to the proposed algorithm, we denote as $\Ust$ the set containing the following configurations: (1) the $\ctype{1}{2}$-configuration, and (2) the $\ctype{2}{2}$-configuration with three occupied vertices. Then, all the configurations in $\Ust $ must be excluded from the possible initial configurations of $\algosquare$.

\subsection{Formalization of the algorithm} 
Algorithm $\algosquare$ processes each configuration not belonging to $\Ust$.
Its strategy can be informally described according to the following steps: first, if necessary, it moves robots so that a corner of $\mbr(C)$ is unoccupied; then, it moves robots to reduce the size of $\mbr(C)$ until reaching a $\ctype{2}{2}$-configuration with two unoccupied corners. Notice that, when a $\ctype{2}{2}$-configuration with two unoccupied corners is obtained, it corresponds to the nice-star configuration as requested by Theorem~\ref{th:star}. Before providing all the formal details, we remark that the algorithm follows the methodology recalled in Section~\ref{sec:prob}. In particular, $\algosquare$ is composed of the following four tasks. 

\begin{description}
    \item[Task 1:] It handles \emph{special cases}, the only cases in which the planned moves enlarge $\mbr(C)$. The precondition $\pre_1$ is true when one of the following two conditions holds: (1) $C$ is a $ \ctype{1}{n}$-configuration with $n\ge 3$, (2) $C$ is a $\ctype{m}{n}$-configuration, $m,n\ge 2$, with all the four corners of $C$ occupied. The corresponding moves are defined as follows. 
    \begin{itemize}
        \item If $C$ is a $\ctype{1}{n}$-configuration, then the active robot  moves to create a $\ctype{2}{n}$-configuration.
        \item If $C$ is a $\ctype{m}{n}$-configuration, $m,n\ge 2$, then a robot moves from the boundary of $\mbr(C)$ to create a $\ctype{m+1}{n}$-configuration.
    \end{itemize}    

    \item[Task 2:] It handles the \emph{general case}, since its precondition $\pre_2$ is true when $C$ is a $\ctype{m}{n}$-configuration, with $m\ge 3$ and $n\ge 2$ but excluding the $\ctype{3}{2}$ case, and with at least one corner unoccupied. Denoting as $v$ an unoccupied corner of $C$, the corresponding move is defined according to the following cases.
    \begin{itemize}
        \item Case $m>n$. Let $L$ be the shortest side of $\mbr(C)$ farthest from the unoccupied corner $v$, and let $r$ be the robot on $L$.  Then, when $r$ is active, it moves, leaving $L$ and entering the region enclosed by $\mbr(C)$.
        \item Case $m=n$. The algorithm works as in the previous case, with the only difference that now $L$ is just any of the two sides of $\mbr(C)$ to which $v$ does not belong. 
    \end{itemize}
    \item[Task 3:] It handles the \emph{pre-final} case, since its precondition $\pre_3$ is true when $C$ is a $\ctype{3}{2}$-configuration with at least one corner of $C$ unoccupied. All the possible managed configurations are reported in Figure~\ref{fig:algo_3x2} along with the move planned by $\algosquare$ in each specific case. The purpose of the task is to form a $\ctype{2}{2}$-configuration with two non-adjacent occupied vertices. 

\begin{figure}[t]
 \centering
 \resizebox{0.8\textwidth}{!}{%
     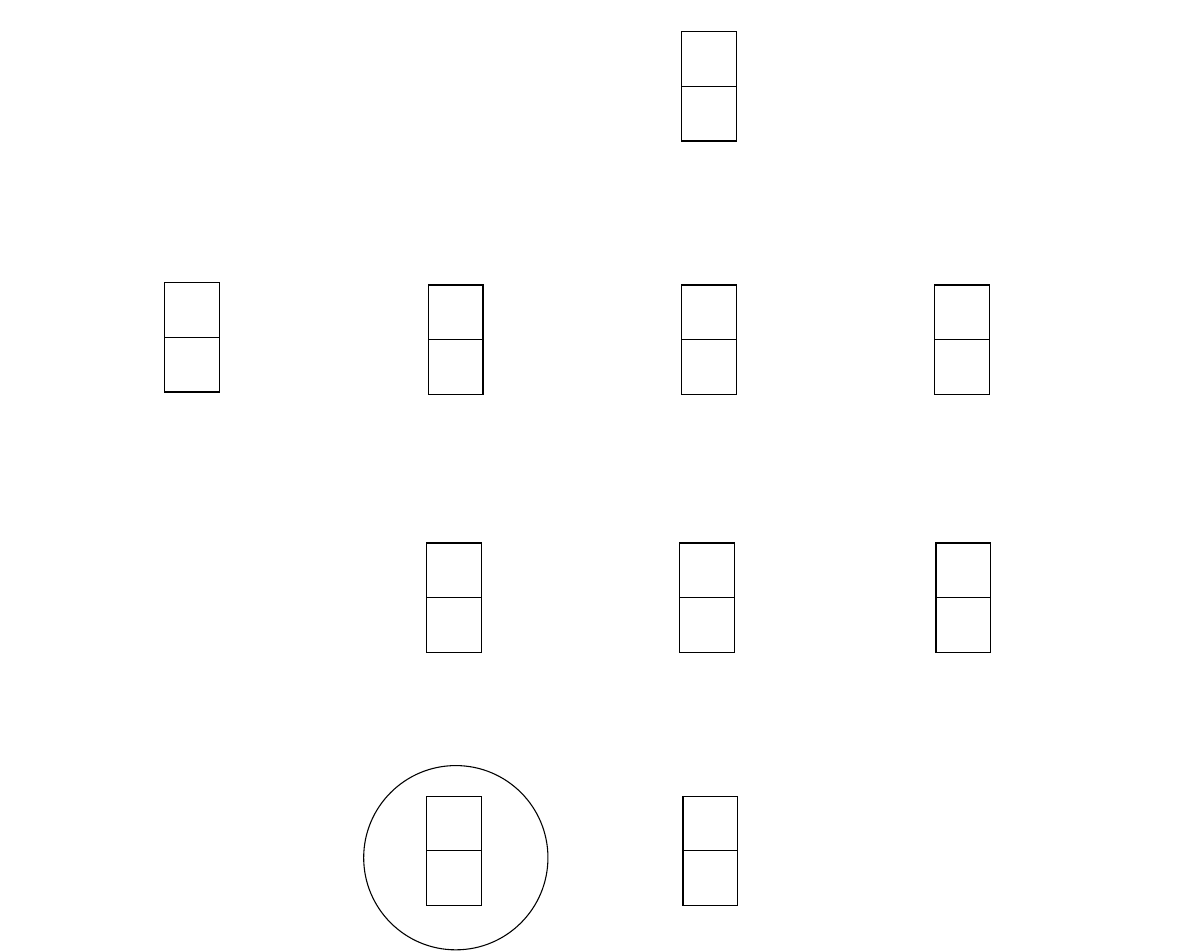
        }
         \caption{The transition graph for all the $\ctype{3}{2}$-configurations. Dashed arrows show moves from vertices occupied by a multiplicity. Dashed vertices highlight the configurations handled by Task $T_4$. For the sake of readability, self-loops are omitted (there is a self-loop in each node where the robot moves toward an occupied vertex).}
\label{fig:algo_3x2}
\end{figure}

    \item[Task 4:] It handles the \emph{final} case, since its precondition $\pre_4$ is true when $C$ is a $\ctype{2}{2}$-configuration with two or three occupied vertices, or a $\ctype{1}{2}$-configuration. The corresponding move is defined according to the following cases:
    \begin{itemize}
        \item $C$ is a $\ctype{2}{2}$-configuration with three occupied corners. A robot moves to form a $\ctype{1}{2}$-configuration. 
        \item $C$ is a $\ctype{2}{2}$-configuration with two occupied corners, and these occupied corners are not adjacent. A robot moves to form a $\ctype{1}{2}$-configuration.         
        \item $C$ is a $\ctype{1}{2}$-configuration. The active robot moves to the other occupied vertex.
    \end{itemize}
    %
\end{description}

%
The following statement proves that $\algosquare$ correctly solves Gathering for each configuration not in $\Ust$.

\begin{table}[t]
    \centering
    \normalsize
    \begin{tabular}{c|l}
    \hline
$T_1$  & $T_2$\\\hline
$T_2$ & $T_2$, $T_3$  \\\hline
$T_3$ & $T_3$, $T_4$\\\hline
$T_4$ & {\gath}\\\hline
    \end{tabular}
    \vspace{0.7em}
    \caption{A tabular representation of the transition graph of $\algosquare$.}
    \label{fig:Ast_transition}
\end{table}
\begin{theorem}\label{teo:square}
Let $G_S$ be a square tessellation graph and let $C=(G_S,\lambda)$ be an initial  $\ctype{m}{n}$-configuration. If $C\not\in \Ust$, then the {\gath} problem can be optimally solved in $C$ under the \rr\ scheduler.
\end{theorem}
\begin{proof}
According to the methodology recalled in Section~\ref{sec:prob}, 
algorithm $\algosquare$ is well defined. 
We need to show that properties $H_1$, $H_2$ and $H_3$ hold. 


In Task 1, $\ctype{1}{n}$-configurations, with $n\ge 3$, are modified into a $\ctype{2}{n}$ type. For $\ctype{m}{n}$-configurations, $m,n\ge 2$ and four occupied corners, the planned move produces a $\ctype{m+1}{n}$ type. In both cases, the configurations produced by this task directly imply that $H_1$, $H_2$ and $H_3$ hold.  

Concerning Task 2, assume $\algosquare$ takes as input a $\ctype{m}{n}$-configuration $C$, with $m\ge 3$, $n\ge 2$ but not of type $\ctype{3}{2}$, and with at least one corner unoccupied. Two different cases must be analyzed.
\begin{itemize}
    \item Assume a corner $v$ of $C$ is unoccupied and $m>n$. In this case, let $L$ be the shortest side of $\mbr(C)$ farthest from $v$, and let $r$ be the robot on $L$. Then, when $r$ is active, it moves, leaving $L$ and entering the region enclosed by $\mbr(C)$. Since $v$ remains unoccupied, this generates a configuration of the same type, unless $r$ is the unique robot on $L$. In such a case, regardless of possible multiplicities, a $\ctype{m-1}{n}$-configuration with an unoccupied corner is generated within one epoch.
    \item If a corner $v$ of $C$ is unoccupied and $m=n$, $\algosquare$ works as in the previous case, with the only difference that now $L$ is just any of the two sides of $\mbr(C)$ which $v$ does not belong to. Also in this case, a $\ctype{m-1}{n}$-configuration with an unoccupied corner is generated in one epoch.
\end{itemize}
According to this case analysis, properties $H_1$, $H_2$ and $H_3$ hold.

Concerning Task $T_3$, the transitions in Figure~\ref{fig:algo_3x2} show that either a new configuration to be handled by the same task is generated, or a $\ctype{2}{2}$-configuration with two non-adjacent occupied vertices is produced and then handled by Task $T_4$. There are no cycles in Figure~\ref{fig:algo_3x2}, except the (omitted) self-loops generated by the robot moving from a vertex with multiplicities toward an already occupied vertex. However, the number of self-loops is limited to the number of robots within the multiplicity, hence, the loop terminates in one epoch. Therefore, the task ends in $O(1)$ epochs. Notice that the reached configuration is the only sink node of Table~\ref{fig:Ast_transition}. In conclusion, even in Task $T_3$, properties $H_1$, $H_2$ and $H_3$ hold.

When Task $T_4$ starts, the handled configuration $C$ is necessarily a $\ctype{2}{2}$-configuration with two occupied vertices and such that those occupied vertices are not adjacent. Notice that in case $T_4$ starts for different types of $\ctype{2}{2}$-configurations, then such configurations have been generated by algorithms starting from more complex initial configurations. By reconsidering the $\ctype{2}{2}$-configuration $C$ with two occupied and non-adjacent vertices, we remark that $C$ represents the requested nice-star configuration (cf Theorem~\ref{th:star}). Here, the algorithm exploits a whole epoch of the \rr\ scheduler: the first robot moving reaches the center of the star (this robot arbitrarily selects the center between the possible twos). This move produces a $\ctype{2}{2}$-configuration with three occupied vertices. When other robots become active in the same epoch, they recognize that such a configuration has been produced by the algorithm itself (in fact, this configuration is never taken as an initial one). Hence, the active robot moves to the occupied vertex, forming the center of the star. Regardless of possible multiplicities, all robots that will be active in the same epoch and observe a $\ctype{2}{2}$-configuration with three occupied vertices will move to the center of the star. This will finally produce a $\ctype{1}{2}$-configuration, handled by the same task. The robots that have yet to move in the same epoch, by moving to the adjacent occupied vertex, will finalize the {\gath}. Notice that, during the execution of the entire task, properties  $H_1$, $H_2$ and $H_3$ hold.
The whole task requires only one epoch. 

Since properties properties $H_1$, $H_2$ and $H_3$ hold in all tasks and there are no loops of the transition graph involving more than one task, the correctness of $\algosquare$ is proved.

Concerning the complexity, the time required by the algorithm is measured in terms of epochs. Recall that $\mbr(C)$ encloses $m$ rows and $n$ columns of the grid. 
Task $T_1$ requires one robot movement, task $T_2$ needs $O(n+m)$ epochs. In the worst case, each robot must move along both the $n$ columns and the $m$ rows. Task $T_3$ requires $O(1)$ epoch, as it reduces a $\ctype{3}{2}$ configuration to a $\ctype{2}{2}$ configuration.
Therefore, the total time required by algorithm $\algosquare$ is $O(n+m)$ epochs. According to Lemma~\ref{lem:lb}, this makes it optimal since $O(n+m)=O(diam(G_S))$. \qed
\end{proof}

\subsection{The characterization} 
We recall that $\Ust$ contains the following configurations: (1) the $\ctype{1}{2}$-configuration, and (2) the $\ctype{2}{2}$-configuration with three occupied vertices. Let us denote such configurations as $C_{\ctype{1}{2}}$ and $C_{\ctype{2}{2}}$, respectively. Moreover, we denote as $C_{\ctype{1}{3}}$ the $\ctype{1}{3}$-configuration with three occupied vertices.



It is not hard to think of a different version of $\algosquare$ (say $\algosquare'$) that reverses the approach by admitting  $C_{\ctype{2}{2}}$ as a possible initial configuration and ignoring $C_{\ctype{1}{3}}$. The strategy of $\algosquare'$, when applied to a generic configuration $C$, could be as follows:
\begin{itemize}
    \item assume that $L$ is a side of $\mbr(C)$ with larger number of occupied vertices (and w.l.o.g., assume $L$ lies on a column of $\mbr(C)$);
    \item if $L$ does not contain one unoccupied vertex, a robot on an endpoint of $L$ is moved to increase the length of $L$ thus obtaining on $L$ one unoccupied vertex;
    \item if $L$ contains one unoccupied vertex,  robots are moved along rows toward $L$ until forming a $\ctype{1}{n}$-configuration $C'$; 
    \item one unoccupied vertex in $\mbr(C)$ is identified to be the center $c$ of a nice-star configuration to be formed;
    \item if the center $c$ is identified, all the robots are accumulated on the two vertices adjacent to $c$ in $\mbr(C)$;
    \item when a nice-star configuration of type $\ctype{1}{3}$ is formed, the last epoch of activations will allow all robots to gather on $c$.
\end{itemize}

The formalization of $\algosquare'$ would lead to the following possible conclusive statement concerning the characterization of Gathering, under the \rr\ scheduler, on square tessellation graphs.

\begin{theorem}\label{teo:square-final} 
Let $C=(G_S,\lambda)$ be an initial configuration where $G_S$ is the infinite square graph. The {\gath} problem can be optimally solved in $C$ under the \rr\ scheduler if and only if the configurations in exactly one of the following sets are excluded from the initial configurations: $\{C_{\ctype{1}{2}}, C_{\ctype{2}{2}}\}$, $\{C_{\ctype{1}{2}}, C_{\ctype{1}{3}}\}$.
\end{theorem}

\section{Concluding remarks and future work}\label{sec:concl}
We have approached the {\gath} problem for a swarm of robots moving on vertex- and edge-transitive graphs under a Round Robin scheduler. 
We first provided some basic impossibility results, and then we designed two different optimal algorithms approaching configurations of robots on infinite grids and hypercubes, respectively.  
Considering also the strategy applied to solve the problem for configurations on rings presented in~\cite{NP25}, we notice that all such algorithms heavily exploit the properties of the underlying topologies. For such a reason, we pose the following conjecture:

\begin{conjecture}
    There not exists a general strategy that can be applied for solving {\gath} under the \rr\ scheduler in vertex- and edge-transitive graphs.
\end{conjecture}

In other words, any strategy that solves {\gath} under the \rr\ scheduler in vertex- and edge-transitive graphs must rely on the specific topology. Perhaps, the strategy applied for infinite square grids can be extended to deal with any tessellation graph, that is, triangular and hexagonal grids.

Further investigations might be conducted on graphs that are vertex-transitive but not edge-transitive, like 
cube-connected cycles $\mathit{CCC}_n$ with $n\ge 3$, or torus grid $C_n\Box C_m$ with $n,m\ge 3$ and $n\neq m$. 






\begin{thebibliography}{10}

\bibitem{BPT16}
Francois Bonnet, Maria Potop{-}Butucaru, and S{\'{e}}bastien Tixeuil.
\newblock Asynchronous gathering in rings with 4 robots.
\newblock In {\em Proc. 5th Int.'l Conf. on Ad-hoc, Mobile, and Wireless
  Networks {(ADHOC-NOW)}}, volume 9724 of {\em LNCS}, pages 311--324. Springer,
  2016.

\bibitem{BKAS18}
Kaustav Bose, Manash~Kumar Kundu, Ranendu Adhikary, and Buddhadeb Sau.
\newblock Optimal gathering by asynchronous oblivious robots in hypercubes.
\newblock In {\em Proc. 20th Int.'l Symp. on Algorithms and Experiments for
  Sensor Systems, Wireless Networks and Distributed Robotics (Algosensors)},
  volume 11410 of {\em LNCS}, pages 102--117, 2019.

\bibitem{CDDN24}
Serafino Cicerone, Alessia {Di Fonso}, Gabriele {Di Stefano}, and Alfredo
  Navarra.
\newblock Gathering of robots in butterfly networks.
\newblock In {\em Proc. 26th Int.'l Symp. on Stabilization, Safety, and
  Security of Distributed Systems ({SSS})}, volume 14931 of {\em Lecture Notes
  in Computer Science}, pages 106--120. Springer, 2024.
\newblock \href {https://doi.org/10.1007/978-3-031-74498-3\_7}
  {\path{doi:10.1007/978-3-031-74498-3\_7}}.

\bibitem{CDDN25-gathering-noVT}
Serafino Cicerone, Alessia {Di Fonso}, Gabriele {Di Stefano}, and Alfredo
  Navarra.
\newblock Gathering in non-vertex-transitive graphs under round robin, 2025.
\newblock URL: \url{https://arxiv.org/abs/2509.06064}, \href
  {https://arxiv.org/abs/2509.06064} {\path{arXiv:2509.06064}}.

\bibitem{SSS_vertex_nonVT_RR}
Serafino Cicerone, Alessia {Di Fonso}, Gabriele {Di Stefano}, and Alfredo
  Navarra.
\newblock Gathering in non-vertex-transitive graphs under round robin.
\newblock In {\em Stabilization, Safety, and Security of Distributed Systems -
  27th International Symposium, {SSS}}, volume 16350 of {\em Lecture Notes in
  Computer Science}. Springer, 2025.

\bibitem{CiceroneFSN25_tcs}
Serafino Cicerone, Alessia {Di Fonso}, Gabriele {Di Stefano}, and Alfredo
  Navarra.
\newblock Optimal gathering of robots in anonymous butterfly networks via
  leader election.
\newblock {\em Theoretical Computer Science}, 1057:115553, 2025.
\newblock \href {https://doi.org/10.1016/J.TCS.2025.115553}
  {\path{doi:10.1016/J.TCS.2025.115553}}.

\bibitem{CDN18-book}
Serafino Cicerone, Gabriele {Di Stefano}, and Alfredo Navarra.
\newblock Asynchronous robots on graphs: Gathering.
\newblock In Paola Flocchini, Giuseppe Prencipe, and Nicola Santoro, editors,
  {\em Distributed Computing by Mobile Entities, Current Research in Moving and
  Computing}, volume 11340 of {\em LNCS}, pages 184--217. Springer, 2019.
\newblock \href {https://doi.org/10.1007/978-3-030-11072-7\_8}
  {\path{doi:10.1007/978-3-030-11072-7\_8}}.

\bibitem{CDN19c}
Serafino Cicerone, Gabriele {Di Stefano}, and Alfredo Navarra.
\newblock On gathering of semi-synchronous robots in graphs.
\newblock In {\em Proc. 21st Int,'l Symp. on Stabilization, Safety, and
  Security of Distributed Systems (SSS)}, volume 11914 of {\em LNCS}, pages
  84--98. Springer, 2019.
\newblock \href {https://doi.org/10.1007/978-3-030-34992-9\_7}
  {\path{doi:10.1007/978-3-030-34992-9\_7}}.

\bibitem{CDN20a}
Serafino Cicerone, Gabriele {Di Stefano}, and Alfredo Navarra.
\newblock Gathering robots in graphs: The central role of synchronicity.
\newblock {\em Theor. Comput. Sci.}, 849:99--120, 2021.
\newblock \href {https://doi.org/10.1016/j.tcs.2020.10.011}
  {\path{doi:10.1016/j.tcs.2020.10.011}}.

\bibitem{CDN21a}
Serafino Cicerone, Gabriele {Di Stefano}, and Alfredo Navarra.
\newblock A structured methodology for designing distributed algorithms for
  mobile entities.
\newblock {\em Information Sciences}, 574:111--132, 2021.
\newblock \href {https://doi.org/10.1016/j.ins.2021.05.043}
  {\path{doi:10.1016/j.ins.2021.05.043}}.

\bibitem{CFPS12}
Mark Cieliebak, Paola Flocchini, Giuseppe Prencipe, and Nicola Santoro.
\newblock Distributed computing by mobile robots: Gathering.
\newblock {\em SIAM J. on Computing}, 41(4):829--879, 2012.

\bibitem{DDKN12}
Gianlorenzo {D'Angelo}, Gabriele {Di Stefano}, Ralf Klasing, and Alfredo
  Navarra.
\newblock Gathering of robots on anonymous grids and trees without multiplicity
  detection.
\newblock {\em Theor. Comput. Sci.}, 610:158--168, 2016.

\bibitem{DDN13}
Gianlorenzo {D'Angelo}, Gabriele {Di Stefano}, and Alfredo Navarra.
\newblock Gathering asynchronous and oblivious robots on basic graph topologies
  under the look-compute-move model.
\newblock In {\em Search Theory: A Game Theoretic Perspective}, pages 197--222.
  Springer, 2013.

\bibitem{DDN14}
Gianlorenzo D'Angelo, Gabriele {Di Stefano}, and Alfredo Navarra.
\newblock Gathering on rings under the look-compute-move model.
\newblock {\em Distributed Computing}, 27(4):255--285, 2014.

\bibitem{DDN11}
Gianlorenzo D'Angelo, Gabriele {Di Stefano}, and Alfredo Navarra.
\newblock Gathering six oblivious robots on anonymous symmetric rings.
\newblock {\em J. Discrete Algorithms}, 26:16--27, 2014.

\bibitem{DDNNS15}
Gianlorenzo {D'Angelo}, Gabriele {Di Stefano}, Alfredo Navarra, Nicolas Nisse,
  and Karol Suchan.
\newblock Computing on rings by oblivious robots: {A} unified approach for
  different tasks.
\newblock {\em Algorithmica}, 72(4):1055--1096, 2015.

\bibitem{DNN17}
Gianlorenzo D'Angelo, Alfredo Navarra, and Nicolas Nisse.
\newblock A unified approach for gathering and exclusive searching on rings
  under weak assumptions.
\newblock {\em Distributed Computing}, 30(1):17--48, 2017.

\bibitem{DDFN18}
Mattia {D'Emidio}, Gabriele {Di Stefano}, Daniele Frigioni, and Alfredo
  Navarra.
\newblock Characterizing the computational power of mobile robots on graphs and
  implications for the euclidean plane.
\newblock {\em Inf. Comput.}, 263:57--74, 2018.

\bibitem{DN17}
Gabriele {Di Stefano} and Alfredo Navarra.
\newblock Gathering of oblivious robots on infinite grids with minimum traveled
  distance.
\newblock {\em Inf. Comput.}, 254:377--391, 2017.

\bibitem{DN17a}
Gabriele {Di Stefano} and Alfredo Navarra.
\newblock Optimal gathering of oblivious robots in anonymous graphs and its
  application on trees and rings.
\newblock {\em Distributed Computing}, 30(2):75--86, 2017.

\bibitem{FPS-macbook19}
Paola Flocchini, Giuseppe Prencipe, and Nicola Santoro, editors.
\newblock {\em Distributed Computing by Mobile Entities, Current Research in
  Moving and Computing}, volume 11340 of {\em Lecture Notes in Computer
  Science}.
\newblock Springer, 2019.
\newblock \href {https://doi.org/10.1007/978-3-030-11072-7}
  {\path{doi:10.1007/978-3-030-11072-7}}.

\bibitem{FPS12}
Paola Flocchini, Giuseppe Prencipe, and Nicola {Santoro (Eds.)}.
\newblock {\em Distributed Computing by Oblivious Mobile Robots}.
\newblock Synthesis Lectures on Distributed Computing Theory. Morgan {\&}
  Claypool Publishers, 2012.

\bibitem{GS87}
B.~Gr\"unbaum and G.~C. Shepard.
\newblock {\em Tiling and Patterns}.
\newblock W. H. Freeman \& Co., New York, 1987.

\bibitem{GP13}
Samuel Guilbault and Andrzej Pelc.
\newblock Gathering asynchronous oblivious agents with local vision in regular
  bipartite graphs.
\newblock {\em Theor. Comput. Sci.}, 509:86--96, 2013.

\bibitem{Ionascu12}
Eugen~J. Ionascu.
\newblock Half domination arrangements in regular and semi-regular tessellation
  type graphs.
\newblock {\em Math}, abs/1201.4624v1, 2012.
\newblock URL: \url{https://arxiv.org/abs/1201.4624v1}.

\bibitem{IIKO13}
Tomoko Izumi, Taisuke Izumi, Sayaka Kamei, and Fukuhito Ooshita.
\newblock Time-optimal gathering algorithm of mobile robots with local weak
  multiplicity detection in rings.
\newblock {\em {IEICE} Transactions}, 96-A(6):1072--1080, 2013.

\bibitem{KLOTW21}
Sayaka Kamei, Anissa Lamani, Fukuhito Ooshita, S{\'{e}}bastien Tixeuil, and
  Koichi Wada.
\newblock Asynchronous gathering in a torus.
\newblock In {\em 25th Int.'l Conf. on Principles of Distributed Systems
  ({OPODIS})}, volume 217 of {\em LIPIcs}, pages 9:1--9:17. Schloss Dagstuhl -
  Leibniz-Zentrum f{\"{u}}r Informatik, 2021.

\bibitem{KKN10}
Ralf Klasing, Adrian Kosowski, and Alfredo Navarra.
\newblock Taking advantage of symmetries: Gathering of many asynchronous
  oblivious robots on a ring.
\newblock {\em Theor. Comput. Sci.}, 411:3235--3246, 2010.

\bibitem{KMP08}
Ralf Klasing, Euripides Markou, and Andrzej Pelc.
\newblock Gathering asynchronous oblivious mobile robots in a ring.
\newblock {\em Theor. Comput. Sci.}, 390:27--39, 2008.

\bibitem{NP25}
Alfredo Navarra and Francesco Piselli.
\newblock Oblivious robots under round robin: Gathering on rings.
\newblock In {\em Proc. 19th Int.'l' Joint Conf. - Frontiers of Algorithmics
  Wisdom ({IJTCS-FAW})}, Lecture Notes in Computer Science. Springer, 2025.

\bibitem{OT12}
Fukuhito Ooshita and S{\'{e}}bastien Tixeuil.
\newblock On the self-stabilization of mobile oblivious robots in uniform
  rings.
\newblock In {\em Proc. 14th Int.'l Symp. on Stabilization, Safety, and
  Security in Distributed Systems (SSS)}, volume 7596 of {\em LNCS}, pages
  49--63. Springer, 2012.

\bibitem{SY99}
Ichiro Suzuki and Masafumi Yamashita.
\newblock Distributed anonymous mobile robots: Formation of geometric patterns.
\newblock {\em {SIAM} J. Comput.}, 28(4):1347--1363, 1999.

\end{thebibliography}
\end{document}